\documentclass[10pt,oneocolumn,preprint]{aastex}
\usepackage{url}

\usepackage{latexsym}
\usepackage{bm}
\usepackage{calrsfs}
\usepackage{graphicx}

\RequirePackage{color}

\newcommand{\emaila}{namouni@obs-nice.fr}

\begin{document}

\title{The excitation of planetary orbits by stellar jet variability and polarity reversal}



\shorttitle{Planetary orbits excitation by stellar jets}
\shortauthors{Namouni}

\author{Fathi Namouni} 
\affil{Universit\'e de Nice, CNRS,  Observatoire de la C\^ote d'Azur, BP 4229, 06304 Nice, France}
\email{\emaila}

\begin{abstract}
Planets form in active protoplanetary disks that sustain stellar jets. Momentum loss from the jet system may excite the planets' orbital eccentricity and inclination (Namouni 2005, AJ 130, 280). Evaluating quantitatively the effects of such excitation requires a realistic modeling of the momentum loss profiles associated with stellar jets. In this work, we model linear momentum loss as a time-variable stochastic process that results in a zero mean stellar acceleration. Momentum loss may involve periodic or random polarity reversals. We characterize orbital excitation as a function of the variability timescale and  identify a novel excitation resonance between a planet's orbital period and the jet's variability timescale where the former equals twice the latter. For constant variability timescales, resonance is efficient for both periodic and random polarity reversals, the latter being stronger than the former. For a time variable variability timescale, resonance crossing is a more efficient excitation mechanism when polarity reversals are periodic. Each polarity reversal type has distinct  features that may help constrain the magnetic history of the star through the observation of its planetary companions.  For instance, outward planet migration to large distances from parent stars is one of the natural outcomes of periodic polarity reversal excitation if resonance crossing is sufficiently slow. Applying the excitation mechanism to the solar system, we find that the planet-jet variability resonance with periodic polarity reversal momentum loss  is a possible origin for the hitherto unexplained inclination of Jupiter's orbit by $6^\circ$ with respect to the Sun's equator. 
\end{abstract}


\section{Introduction}
The discovery of more than 760 exoplanets to date \citep{b18} has revolutionized our understanding of the architecture of planetary systems. One of the surprising key observations is the large eccentricity of exoplanet orbits. Compared to Jupiter's eccentricity of 0.05 and to the Earth's 0.017, the median eccentricity of exoplanet orbits is 0.2. Various dynamical mechanisms were proposed to account for this departure from the solar system's planetary standard. These mechanisms include: planet-planet scattering \citep{scatt3,scatt1,scatt2},  three-body secular Kozai oscillations \citep{koz2,koz3,koz1}, mean motion resonances \citep{res1,res2,res3}, stellar encounters \citep{stenc} and excitation induced by stellar jets (\cite{b19} hereafter Paper I, \cite{rj8}). Although some mechanisms are more efficient than others, attention has  recently been devoted to planet-planet scattering owing to the claim that this mechanism reproduces the observed eccentricity distribution of exoplanet orbits. However, the eccentricity distribution produced from planet-planet scattering only approximates the observed distribution if the latter is cut off at an eccentricity of 0.2  \citep{scatt2}. Smaller planetary orbital eccentricities were attributed to the other possible mechanisms or a combination thereof. The threefold increase in the number of planets between 2008 and 2012 has provided better statistics and decreased the median eccentricity from 0.3 to 0.2 (or 0.23) if only planets with periods larger than 5 (or 20) days are included to account for tidal evolution bias.   The excess of planets in the current enlarged sample with eccentricities smaller than 0.3 with respect to the Rayleigh distribution produced by planet-planet scattering \citep{scatt2} requires that the eccentricity cutoff applied to the observed distribution be increased to 0.3 leaving more than half the planet population with stirred orbits outside the scope of applicability of planet-planet scattering.

In this paper, we revisit the excitation of planetary orbits that results from momentum loss through stellar jets (Paper I) and attempt to assess quantitatively the effects of this mechanism. Orbital excitation through stellar jets is based on jet-counterjet asymmetry  that has been observed in a significant fraction of star-disk systems \citep{b10,b11,b12,b14,b13,hh7, rj1,rj2, rj3, rj4, rj5}. As the planet orbits around the star and the inner disk, it sees that system accelerating away from it owing to asymmetric momentum loss. Excitation from a smooth time-varying jet-induced acceleration is a secular process that requires that the jet axis be inclined with respect to the disk's plane. It was shown in Paper I that the maximum eccentricity achieved is proportional to the sine of the mutual inclination of the planetary orbital normal and the jet axis. In this paper, we develop a realistic modeling of momentum loss as a time-variable stochastic process that results in a zero mean stellar acceleration and restrict our attention to systems where the jet system's axis is perpendicular to the initial planetary plane and secular excitation is absent. Momentum loss models include periodic or random polarity reversals that may be associated with the magnetic polarity reversals of the parent star and the inner star-disk interface. In section 2, we recall the basics of orbital excitation through stellar jets. In section 3, we characterize orbital excitation as a function of the variability time scale and acceleration standard deviation as well as in the presence of mutual planetary perturbations. We identify a fundamental excitation resonance between the planet orbital period and the variability timescale and show that it is an efficient excitation mechanism for both periodic and random reversal jet profiles at constant variability timescale.  In section 4, we examine resonance crossing by modeling the time dependence of the variability time scale. In particular, it is found that resonance crossing  is an efficient excitation mechanism for periodic polarity reversal profiles but not for random polarity reversal profiles.  In the solar system, we find that resonance crossing with periodic polarity reversal and a time variability timescale that increases from 0.5 to 11 years (the current solar cycle half period) is able to reproduce Jupiter's and Saturn's orbital configuration and their inclination by $6^\circ$ with respect to the solar equator. Section 5 contains concluding remarks.

\section{Orbital excitation by asymmetric jet momentum loss}

Stellar jet asymmetry is observed in a increasing number of systems  as the ejection velocities of the jet and counterjet differ by about a factor of 2 \citep{b10,b11,b12,b14,b13,hh7, rj1,rj2, rj3, rj4, rj5}. Jet launching regions (JLRs) are confined to the inner part of the disk with estimates from 0.01 AU for the X-wind model to a few AU for disk-wind models (e.g. 1.6 AU for RW Auriga, \citealt{b15}). The integrated momentum loss over the launching region accelerates the center of mass of the star-disk system which coincides with the star's center for axisymmetric star-disk systems. Gauss's theorem for the gravitational potential of the form $r^{-1}$ implies that as the planet orbits around the star and the inner disk, it sees that system accelerating away from it. The planet's orbital evolution may be described by the equation (\cite{rj7} hereafter Paper II,  \cite{rj6}):
\begin{equation}
\frac{{\rm d} {\bm v}}{{\rm d} t}=-\frac{GM}{|{\bm x}|^3}\,{\bm
  x} +{\bm A}, \label{motion}
\end{equation}
where ${\bm x}$ and ${\bm v}$ are the relative position and velocity of the planet with respect to the star and $G$ is the
gravitational constant.  The mass $M$ is that of the star augmented by that of the inner disk  and corresponds to the total mass contained in the JLR. Away from the JLR, the acceleration ${\bm A}$  that results from momentum loss is not a function of the distance to the star  as it is the ratio of the integrated momentum loss over the jet launching region to the total mass $M$.  If planets enter the JLR, they will be subject to an acceleration that is a monotonically increasing function of the distance to the star reaching the constant value $A$ at the outer edge of the JLR. Such a situation does not concern us in the present work. The effect of the disk's mass and its variations in $M$ may be neglected because the mass of the inner disk is small compared to the mass of the star and also because the mass loss is small and amounts to a correspondingly small dynamical effect given by the Jeans radial migration rate $\dot
r/r=10^{-8}$\,yr$^{-1}$ for $\dot M=10^{-8} M_\odot$\,yr$^{-1}$ \citep{jeans}. An order of magnitude estimate of the acceleration is given by:
\begin{equation}
A \sim  10^{-13}\, 
\left(\frac{\dot M}{10^{-8} M_\odot\, \mbox{\rm  yr}^{-1}}\right)\, 
\left(\frac{v_e}{300 \,\mbox{\rm km\,s}^{-1}}\right)\,
\left(\frac{M_\odot}{M}\right) \mbox{\rm km\,s}^{-2}. \label{acc-mag}
\end{equation}
where $\dot M$ is the mass loss rate, and $v_e$ is the jet's ejection velocity \citep{b9}. This estimate is an instantaneous lower bound on the total momentum loss as we lack long time span observations of stellar jets as well as observations of the jet engine within a few AU from the star. Although small, the acceleration amplitude (\ref{acc-mag}) was shown to be a
possible origin for the large eccentricities and secular resonances of extrasolar planets' orbits provided the planets formed in a jet-sustaining disk (Paper I). The planet's orbital excitation time associated with the acceleration $A$ is  given as $T_A=GM T/3a^2|A|$ where $T$ and $a$ are the orbital period and semimajor axis of the planet. Where $T_A\gg T$, orbital excitation is adiabatic and the eccentricity secular increase is given as: $e(t)= |\sin(\pi t/T_A)\ \sin I_0|$ where $t$ is time and $I_0$ is the inclination of the jets' axis to the planet's initial orbital normal (Paper I). This shows that eccentricity excitation does not occur when $I_0$ vanishes. In deriving this eccentricity evolution expression, acceleration was assumed to be constant when the jets are active. When jet-induced acceleration increases as during the initial launching of the jets or decreases in their final stages of activity, it was shown that the part of the disk outside the jet launching region expands, contracts  and heats up as its state of minimum energy deviates from the usual star-disk mid-plane to a sombrero-shaped profile curved along the direction of acceleration (Paper II). 

In the next section, we will find out under what circumstances it is possible for a jet system that is orthogonal to the planetary orbits (i.e. $I_0=0$) to excite them significantly. By modeling realistic time-variable jets (in contrast to the smooth models of Papers I and II), we show that there is a resonance between the variability timescale and the planet orbital period that may lead to large eccentricity and inclination as well as significant outward radial migration. 

\section{Planet-jet variability resonance}

Stellar jets are time-variable processes \citep{var1, var4,var2, var3, b7,varn2,varn1,varn3,varn4,varn5}. Variability is attributed to the variations of the ejection velocities of the jet and counter jet possibly from mass loading at the base of the jets or from a variable magnetic field of the star and the star-disk interface. 
As we lack long duration observations of stellar jets (the first such observations were made in 1994), modeling realistic asymmetric momentum loss from the jet system requires some assumptions.  We choose the jet axis to be orthogonal to the initial planetary orbital plane. We model the time variations of the asymmetric momentum loss as a stochastic process. The acceleration that results from  momentum loss is drawn from a normal distribution with zero mean and finite standard deviation $\sigma_A$ after a time $\tau$ that we call the variability time scale. Acceleration remains constant for the duration $\tau$ \footnote{If a planet enters the JLR, then $\sigma_A$ will depend on the star-planet distance. This situation does not occur in this work owing to the small accelerations involved.}. The mean acceleration is set to zero because we assume that the observed asymmetry in stellar jets is equally time variable in its direction. It is indeed reasonable to assume that an excess of linear momentum loss from one side of the accretion disk may in time be reversed by loss from the other side especially if the asymmetry's origin  is related to the magnetic field configuration which may be subject to polarity reversals such as those observed in the solar cycle and other stars \citep{mc2,mc6,mc7, mc3,mc4,mc1, mc5,mc8}. 

We consider two possible time variations for the momentum loss process. Acceleration may reverse periodically with a period equal to the variability time scale $\tau$ or it may reverse randomly. Periodic reversals acceleration profiles are constructed from a normal distribution that is forced to reverse with the period $\tau$. For random reversals, the latter forcing is turned off.  The equations of motion (\ref{motion}) are integrated for a given acceleration profile. For all simulations we  monitor the residual  velocity $V$ that the star acquires at the end of simulation. We use enough profiles for a given parameter set so that the mean residual  velocity and its standard deviation are less than a few km\,s$^{-1}$ which is much smaller than the stellar velocity dispersion in the Galaxy of order  $\sim 30$km\,s$^{-1}$ \citep{r5}.

When quoting the acceleration standard deviation, $\sigma_A$, we prefer to use the more physical keplerian boundary semimajor axis $a_{\rm kplr}$ defined as the distance from the star where its  gravitational pull balances the secular excitation by the acceleration $\sigma_A$ (Paper I, \cite{rj6}). Beyond this distance, orbits are no longer bound to the star. This distance is obtained by equating the excitation time $T_{\sigma_A}$ to the local orbital period $T$ and is given as:
\begin{equation}
a_{\rm kplr}=\left(\frac{GM}{3\sigma_A}\right)^\frac{1}{2}\simeq 10^3\,
\left(\frac{2\times 10^{-12}\ {\rm km}\, {\rm s}^{-2}}{\sigma_A}\right)^\frac{1}{2}\, \left(\frac{M}{M_\odot}\right)^\frac{1}{2}\, {\rm AU}.\label{akplr}
\end{equation}

Figures 1 and 2 show the results of two simulations of a Jupiter mass planet located 5 AU away from a solar mass star subject to an acceleration from a variable jet system with an acceleration standard deviation $\sigma_A$ corresponding to $a_{\rm kplr}=200$ AU and a variability time scale $\tau=6$ years. Figures 1 and 2 correspond to periodic and random acceleration reversals respectively (see the acceleration profile panels therein). The excitation of inclination, $I$,  is found to be more important than that of eccentricity, $e$, and radial migration $\Delta a$.  The reason is as follows: although acceleration is perpendicular to the planet's orbit, the orbit does not at any given time lie in the surface of least energy of the accelerated star. It was shown in Paper II that in the presence of acceleration least energy orbits remain circular but their orbital plane no longer includes the star. These orbits hover above the star's equator and lie on a sombrero-shaped surface that is curved along the direction of acceleration. As the sombrero profile varies stochastically with the acceleration, inclination is systematically excited. Only through the conservation of the vertical component of angular momentum that the orbital semimajor axis and eccentricity are forced to change. It was shown that these changes are of second order with respect to the perturbation. Except for the early orbital excitation phase, little difference is seen between the outcomes of periodic and random acceleration. Figures 1 and 2 however show each only one realization of a stochastic process. We estimate the number of runs required to reach a significant conclusion by simulating the planet's evolution over $10^6$ years with 1000 periodic reversal acceleration profiles with the same variability timescale and acceleration standard deviation. Figure 3 shows the dependence of the mean values and standard deviations of radial migration, eccentricity, inclination and residual velocity $V$ of the system in Figure 1 on the number of simulations. It is seen that more than 100 simulations are required to ascertain the probable outcome of a given parameter set. In particular, the average inclination is about $5^\circ$ twice as low as that of Figure 1. We therefore choose for a given parameter set to integrate 500 acceleration profiles in order to determine the excitation outcome.  A similar conclusion was reached for random reversal simulations.

We now address the dependence of excitation on the variability time scale. Figure 4 shows the orbital elements and residual velocities for the reference Jupiter planet at 5 AU after $10^6$ years orbiting the sun subject to a periodic reversal acceleration with a standard deviation of $a_{\rm kplr}=200$ AU. Orbital excitation extrema are found near the resonance of the variability timescale, $\tau$,  with half the orbital period, $T/2$. Maxima occur at half integer periods while minima occur at integer periods. For instance, if the current solar cycle of 22 years (of inversion timescale $\tau=11$\,years) had influenced the polarity reversal of the early solar system's jets then Jupiter at 5.2\,AU (period 11 years) would have been sitting near a location of minimum excitation. As the solar cycle period is likely to have evolved from a smaller period than 22 years, resonance crossing is likely to have  excited the planet's orbit. This point is detailed in the next section. We note that whereas eccentricity peaks at resonance $\tau=5.6$ years, inclination and semi major migration peak slightly before it. Also unlike eccentricity excitation maxima which are periodic with respect to the variability timescale, those of radial  migration and inclination excitation decrease significantly with increasing variability timescale. Figure 5 shows how excitation evolves in time at $10^3$, $10^4$, $10^5$ and $10^6$ years indicating is power-law type dependence. We estimate this dependence as follows. The maximum inclination excitation during a duration $\tau$ when the acceleration is constant is obtained  by solving the vertical equation of motion in
(\ref{motion}) under the assumption that $\sin I\ll 1$ where $I$ denotes inclination. It is found that $z={Aa^3}/{GM}\ (1-\cos n t)\label{vertical}$ and therefore the average inclination increment $\delta I= (a/a_{\rm kplr})^2/3$. Assuming excitation to be a random walk in inclination with increments $\delta I$ that take place $t/\tau$ times, we find an inclination amplitude of:
\begin{equation}
I={f(\tau)}\left(\frac{t}{\tau}\right)^\frac{1}{2} \left(\frac{a}{a_{\rm kplr}}\right)^2 \label{scaling}
\end{equation}
where $f$ is a function that depends on the variability timescale $\tau$. We checked that this scaling has the right dependence on time and also on $a_{\rm kplr}$ by simulating a smaller acceleration corresponding to 300 AU (Figure 6). In Paper II, we showed that the conservation of the vertical component of angular momentum that forces the mean orbital radius to change in order to
compensate for the increase of the vertical motion $\delta I $ leads to a radial migration $\delta a$ and eccentricity increments $\delta e$ proportional to $(a/a_{\rm kplr})^4$. We checked that these scalings apply to the final excitation amplitudes of eccentricity and radial migration. Random polarity reversal simulations have similar scalings as but their amplitudes (e.g. $f(\tau)$ in Equation (\ref{scaling})) are much larger than those of periodic polarity reversal (Figure 7). This difference is due to the fact that whereas $\tau$ is both the acceleration amplitude variability timescale and the polarity reversal timescale for periodic reversal simulations, it is only the minimum polarity reversal timescale and the acceleration amplitude variability timescale for random reversal simulations. For the latter simulations, acceleration is on the same side of the disk plane for a longer time than $\tau$ and can excite the planets more strongly than in the corresponding case with periodic reversal.

Mutual planet perturbations are known to affect orbital excitation by asymmetric momentum loss. In the case of the secular excitation regime studied in  Paper I, mutual gravitational interaction between two planets works against excitation by asymmetric momentum loss through secular perturbations of the orbits. If the jet-induced acceleration is not strong enough to impose pericenter alignment during excitation, the eccentricity and inclination amplitudes may be reduced significantly with respect to their values obtained without mutual perturbation. With stochastic excitation the situation is different and the process of pericentre alignment is irrelevant as there is no preferred excitation direction as far as the planet's orbit is concerned. In effect, the inclination and eccentricity increments acquired at each acceleration amplitude stochastic change redefine a new excitation direction for the orbit that is consequently random. We therefore expect that mutual planet perturbations to reinforce the resonant excitation by stochastic asymmetric momentum loss because the minimum excitation amplitude of the inner planet will not correspond to an excitation minimum for the outer planet. Entrainment by mutual interactions may therefore increase the excitation amplitude of the inner planet. Figure 8 shows the results of the interaction of a Jupiter-mass planet at 5\,AU and a Saturn-mass planet at 8.5\,AU under the influence of periodic polarity reversal momentum loss  with an acceleration standard deviation of $a_{\rm kplr}=200\,$AU. Each point corresponds to 500 acceleration profiles each  integrated over $10^6$ years. The outer planet was not set at 9\, AU in anticipation of the outward migration that orbital excitation will generate. The excitation minima of eccentricity and inclination (but not the semi major axis) for the inner planet are erased. The amplitudes dependence follows to a certain extent the envelope of both inner and outer resonances respectively at $\tau=5.6$ and $12.4$ years  (based on the initial semi major axes). We also note that the orbits' relative inclination ($\sim 2^\circ$--$4^\circ$) is quite large compared to that of Jupiter and Saturn in the solar system ($\sim1^\circ$) whereas their mean inclination ($\sim 5^\circ$--$7^\circ$) is near the observed value of $6^\circ$. Applying a random polarity reversal stochastic acceleration to the Jupiter-Saturn pair will result in larger excitation amplitudes (as seen in Figure 7) but with a similar dependence on the variability timescale. For conciseness, we do not report  the corresponding figures and move to the more physically relevant phenomenon of resonance crossing where the variability timescale is itself time-variable.

\section{Time-dependent variability timescales and resonance crossing}
The identification of the planet-jet variability resonance using a constant variability timescale allowed us to characterize quantitatively the excitation amplitudes of the eccentricity, inclination, and semi major axis migration. In reality however, it is likely that there are several variability timescales associated with momentum loss some of which may even vary with time. This would occur particularly if the variability timescale is related to the polarity reversals of the magnetic field of the star-disk interface within the jet-launching region. Stellar polarity reversal cycles are observed in several stars and seem to be correlated with the rotation rate as solar-mass stars rotating faster than the Sun tend to have shorter polarity reversal times \citep{mc5}. For instance, spectropolarimetric observations using Zeeman Doppler imaging showed that the planet hosting star $\tau$~Bootis has cyclic polarity reversals with $\tau=1$ year \citep{mc2} although such a cycle has yet to be confirmed by X-ray observations and optical spectra of the star  \citep{mc8}. Numerical simulations of the generation of magnetic fields in young rotating stars confirm the star rotation-magnetic cycle period correlation and showed that a solar mass star rotating five times faster than the Sun had polarity reversals with $\tau=4$ years instead of the solar value $\tau=11$ years \citep{mc1}. As a star evolves toward slower rotation and hence toward larger magnetic cycle periods, the surrounding planets will cross the corresponding planet-jet variability resonances. We assess the effect of resonance crossing on orbital excitation by modeling the evolution of a Jupiter-mass planet at 5 AU under stochastic momentum loss whose variability time scale $\tau$ increases from $\tau=0.5$ to 11 years. As the outcome of resonance crossing depends on how fast resonance is traversed, we use the time it takes $\tau$ to reach 11 years, denoted $t_r$, as our parameter of crossing velocity. For the time dependence of $\tau$ we experimented with various exponential and power-type laws.  Here we report on two time profiles. The first is the square root law that comes up naturally if the variability timescale is assumed to increase at a fixed rate $\Delta \tau$ after a duration equal to $\tau$. In this case $2\tau=[(\Delta \tau+ 2\tau_i)^2+ 8 \Delta\tau t ]^{1/2}-\Delta \tau$ where $\Delta \tau=(\tau_f^2-\tau_i^2)/2t_r$ and $\tau_i=0.5$ years and $\tau_f=11$ years are the initial and final values.  The exponential profile is given as $\tau= \tau_f-(\tau_f-\tau_i)\exp(-5t/t_r)$. For a time variable timescale, convergence of the stochastic simulations towards a mean, for a given acceleration parameter set, is achieved after a few hundred runs. We therefore increased the number of simulations to 2000 each integrated for the duration $t_r$.  Figure 9 shows the amplitudes obtained from resonance crossing for a single Jupiter mass planet as a function of the duration $t_r$ (we remind the reader that each curve point corresponds to a different time evolution of the variability time scale $\tau$). Unlike the case of constant variability time scales, it is periodic and not random polarity reversal that achieves the strongest excitation amplitudes (compare upper and lower row amplitudes). Maximum inclination for random reversal simulations is even smaller than those with constant variability timescales (Figure 7). For periodic polarity reversal simulations, resonance crossing is evident through two features: the first is  the ``sudden" increase of eccentricity and inclination amplitudes as well as outward semi major axis migration. This shows that fast crossing of the planet-jet variability resonance is an inefficient mechanism to excite planet orbits --for the square root (exponential) law, resonance crossing occurs around $0.25 \,t_r$ ($0.14\,t_r$).  The second feature is the large excitation amplitudes compared to the corresponding values for constant variability timescales (Figures 4 and 5). Maximum inclination for a constant variability timescale with  $a_{\rm kplr}=200$ AU is of order $2^\circ$ and $5^\circ$  after $10^5$ and  $10^6$ years respectively whereas it peaks at $25^\circ$ for a time variable $\tau$. Acceleration strength affects resonance crossing mainly through the onset of excitation as the excitation amplitudes remain more or less comparable. For the square root law,  strong excitation by resonance crossing  is triggered  after $t_r=1.2\times 10^5$ years for $a_{\rm kplr}=200$ AU and is delayed until  $t_r=3.7 \times 10^5$ years for $a_{\rm kplr}=300$ AU.
The choice of a relaxation time of $t_r/5$ in the exponential law makes resonance crossing inherently faster than that of the square root law explaining why strong excitation is triggered only after $t_r=3 \times 10^5$ for the same acceleration strength ($a_{\rm kplr}=200$ AU). In effect, the crossing velocity at resonance for the exponential law  reads $d\tau/dt=2.5 \, \tau_f/t_r$ whereas it is given as $d\tau/dt=\, \tau_f/t_r$ for  the square root law. The resonance crossing velocity that triggers excitation for the power law at $t_r= 1.2\times 10^5$ is $d\tau/dt = 10^{-4}$ and corresponds to the observed longer time  $t_r$ for the exponential law. The exponential law however has different excitation amplitudes especially that of eccentricity. As the crossing velocity becomes smaller with increasing $t_r$, inclination excitation and orbital migration decrease slowly from their peak values whereas eccentricity excitation becomes independent of crossing velocity. Going back to the random polarity reversal simulations, we note that excitation amplitudes increase monotonically with a decreasing crossing velocity (increasing $t_r$) and that unlike periodic polarity reversal, excitation is weaker for the exponential law. 

We assess the effect of mutual planet interactions on orbital excitation using the two-planet model of the previous section. Figure 10 shows the excitation amplitudes of Jupiter and Saturn that were initially on circular orbits at 5 AU and 8.5 AU respectively. As with the single planet simulations, random polarity reversal leads to smaller excitation amplitudes. For periodic polarity reversal simulations, mutual planet interactions modify the excitation amplitudes of the inner planet as maximum inclination is reduced by $\sim 40\%$ whereas maximum eccentricity is increased by $\sim 50\%$ regardless of acceleration strength and timescale time-dependence law. The outer planet has an interesting response to stochastic excitation in that: first, its inclination excitation profile and amplitude are similar to those of the inner planet alone (Figure 9). Second, its eccentricity excitation profile is an order of magnitude larger than that of constant timescale simulations (Figure 8).  Third, for slow resonance crossing velocities, the planet may migrate several hundreds of AU away from the star. For periodic polarity reversal simulations, the mean relative inclination of the two planets is small before resonance  crossing is able to strongly excite the orbits as well as slightly afterwards. For smaller resonance crossing velocities $d\tau/dt$, relative inclination  becomes quite as large as the excited inclination of the inner planet with respect to the star's equator, and  correlated with significant outward migration. For random polarity reversal, relative inclination increases steadily with decreasing crossing velocity whereas substantial outward migration is absent.

We can use the excitation amplitudes of Figure 10 to locate the actual orbits of Jupiter and Saturn  and constrain the momentum loss process in the early solar system. The relative inclination of the two giant planets that is less than $1^\circ$ disfavors  random polarity reversal momentum loss as small relative inclinations are correlated with much smaller eccentricities than those observed. We are left with periodic polarity reversal momentum loss. To determine the resonance crossing velocity or equivalently $t_r$,  we combine the small relative inclination of Jupiter and Saturn with the observed inclination of Jupiter's orbit with respect to the Sun's equator (6$^\circ$) that essentially defines the invariable plane of the solar system as Jupiter is its most massive body.  The current orbits would then be obtained shortly after the onset of resonance crossing excitation with the power law near $t_r=1.2$ to $1.3\times 10^5$ years and $a_{\rm kplr}=200$ AU. The semi major axis migration of the planets brings their initially smaller orbits to their current sizes. Choosing a weaker acceleration standard deviation with $a_{\rm kplr}=300$ AU and $t_r= 3.5 \times 10^5$ years would produce a larger semi major with respect to Saturn's current orbit only because the planet was started at 8.5 AU. In this respect, our results about the solar system should be regarded as a demonstration of principle of how stochastic momentum loss can produce the current orbits of Jupiter and Saturn. In this sense, this demonstration is quite encouraging if we recall that each amplitude shown in Figure 10 is an average over 2000 momentum loss profiles. In order to constrain more precisely the momentum loss process, all planets need to be included as well as minor bodies and in particular those that are decoupled dynamically from the solar system's planets such as dwarf planet Sedna that may be accounted for by the significant outward migration from smaller orbits produced by stochastic momentum loss. 

 \section{Conclusion}
 In this work we examined quantitatively the excitation of planetary orbits by stellar jet stochastic momentum loss that on average does not accelerate the star. We modeled momentum loss using two main parameters, the acceleration standard deviation and the variability timescale, along with two polarity reversal modes, random and periodic. In particular, we did not invoke a prior inclination of the jet axis as it was taken to be perpendicular to the initial planetary orbits. Whereas secular excitation by asymmetric momentum loss  requires such inclination (Paper I), stochastic momentum loss does not and may achieve far greater amplitudes than secular excitation. Stochastic momentum loss  is efficient at the resonance of the planet's period with the variability timescale. Random polarity reversal appears to cause greater excitation for constant variability timescales but it fails to compete with periodic polarity reversal when the variability timescale is time dependent. If polarity reversal is related to the magnetic field of the star-disk interface, then the reversal timescale will increase during the braking of the star's rotation as indicated by observations of solar-type stars and numerical simulations of young stars' magnetic fields. As the variability timescale increases resonance crossing by the planets' orbits may excite them significantly. We have characterized such excitation and showed that the greater diversity of orbital outcomes occurs with periodic polarity reversal and is determined by how fast the planet-jet variability resonance is crossed. The smallest crossing velocities produce the most extreme systems. In particular, planets can migrate  several hundred AU away from the planet forming region around a solar mass star. Periodic polarity reversal stochastic momentum loss can statistically explain the current configuration of the solar system's Jupiter and Saturn and particularly the hitherto unknown origin of the inclination of Jupiter's orbit with respect to the solar equator. Although our study was focused on planetary systems, stellar companions of a jet-sustaining star are affected similarly by stochastic momentum loss. Perhaps the most the promising result in this work is the possible link between stellar magnetic cycles and the dynamical architecture of planetary companions. This may prove a valuable tool to constrain the magnetic history of planet-hosting or binary stars and understand the observed diversity of planetary systems.

 \acknowledgements
 The author thanks the reviewer for useful comments. The numerical calculations in this work were done at the high performance computing center M\'esocentre {\sc sigamm} hosted at the Observatoire de la C\^ote d'Azur.

\newpage

\begin{figure}
\begin{center}
\includegraphics[width=75mm]{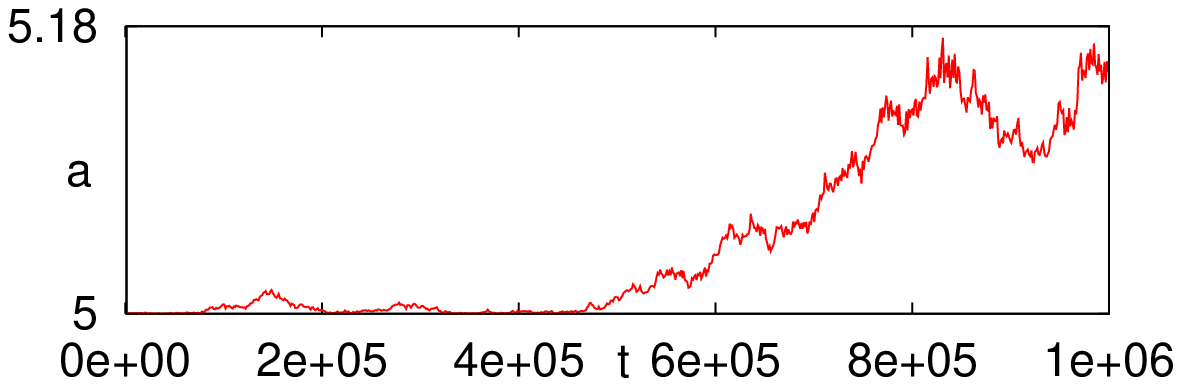}
\includegraphics[width=75mm]{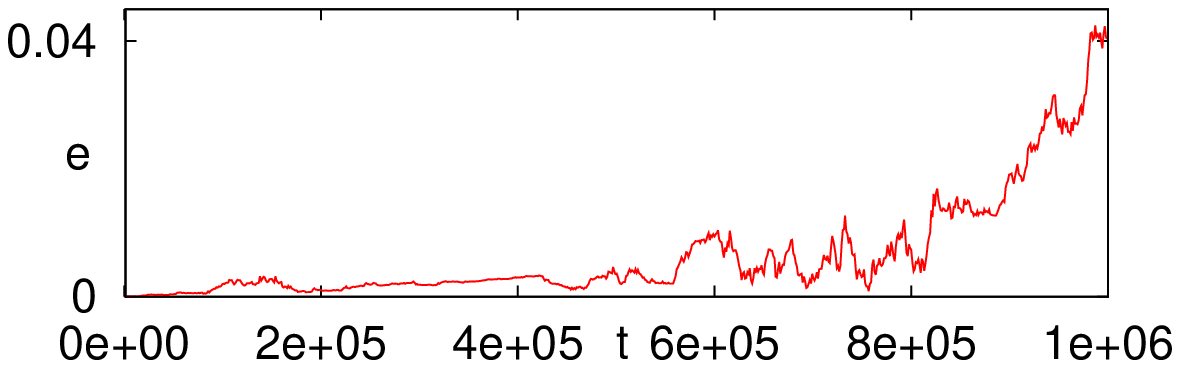}\\
\includegraphics[width=75mm]{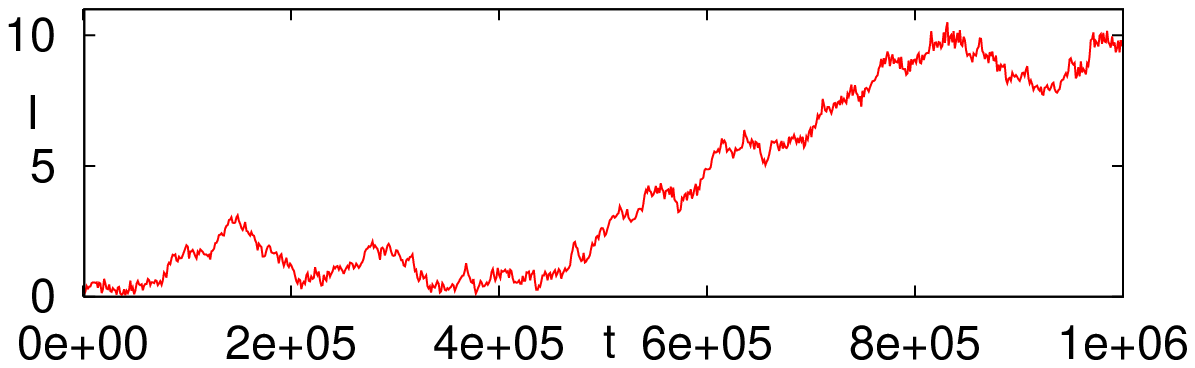}
\includegraphics[width=75mm]{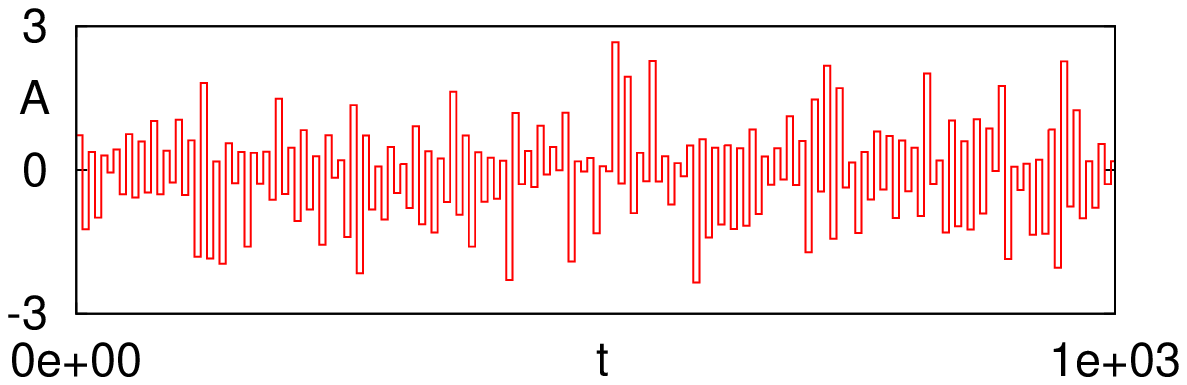}
\end{center}
\label{f1}
\caption{Excitation of a planetary orbit by a periodic polarity reversal variable stellar jet system. The Jupiter mass planet is located at 5 AU and is excited by an acceleration with $a_{\rm kplr}=200$ AU (standard deviation) and a variability timescale  $\tau=6$ years. Shown are the semimajor axis $a$(AU), eccentricity $e$, inclination, $I(^\circ)$ and acceleration $A$ in units of $\sigma_A$ as functions of time (years).  } 
\end{figure}
\newpage
\begin{figure}
\begin{center}
\includegraphics[width=75mm]{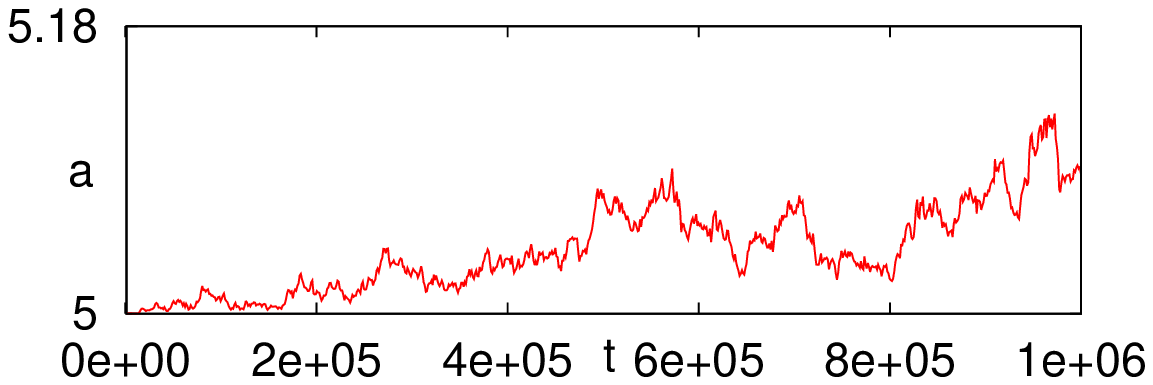}
\includegraphics[width=75mm]{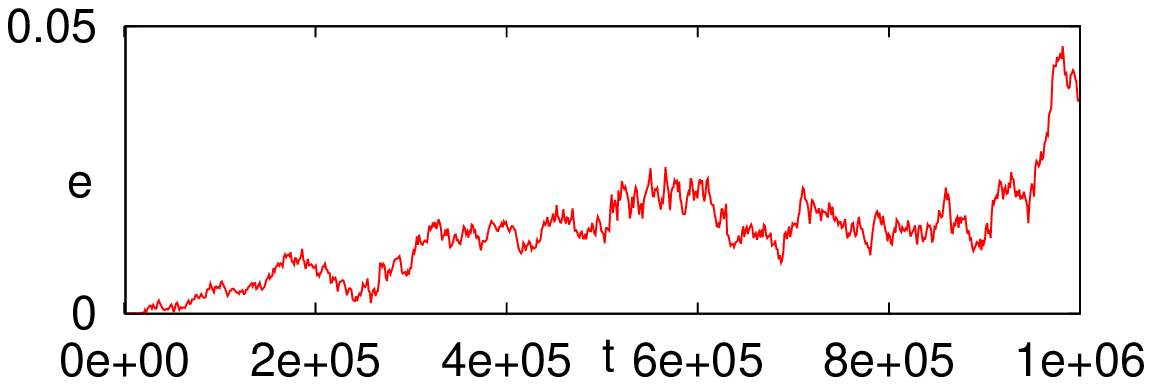}\\
\includegraphics[width=75mm]{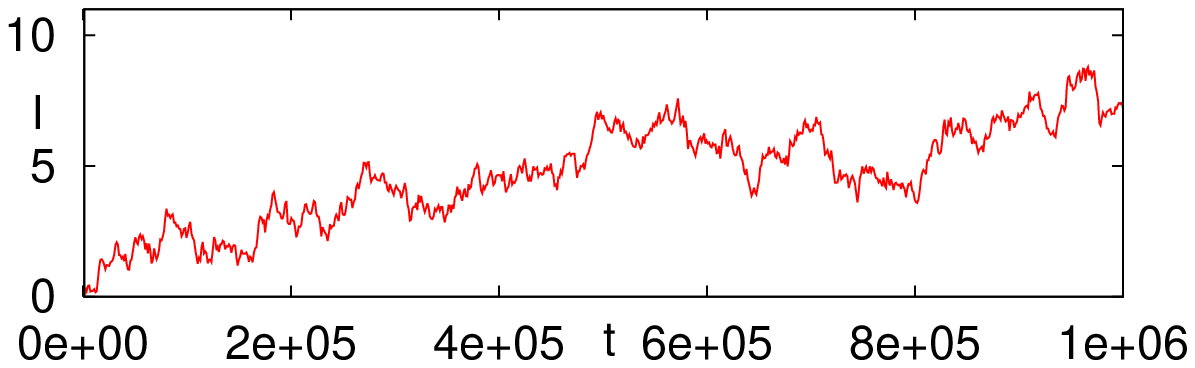}
\includegraphics[width=75mm]{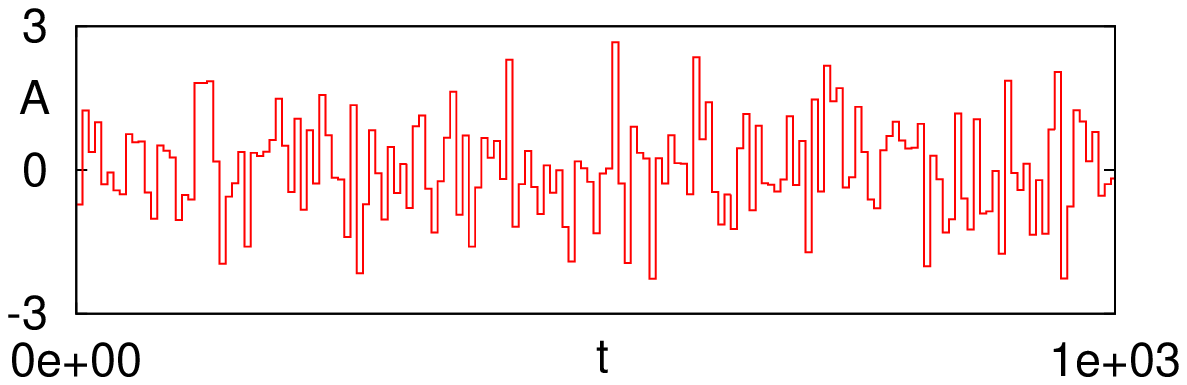}
\end{center}
\label{f2}
\caption{Excitation of a planetary orbit by a random polarity reversal variable stellar jet system. The parameters are those of Figure 1 except for polarity reversal.} 
\end{figure}
\newpage

\begin{figure}
\begin{center}
\includegraphics[width=75mm]{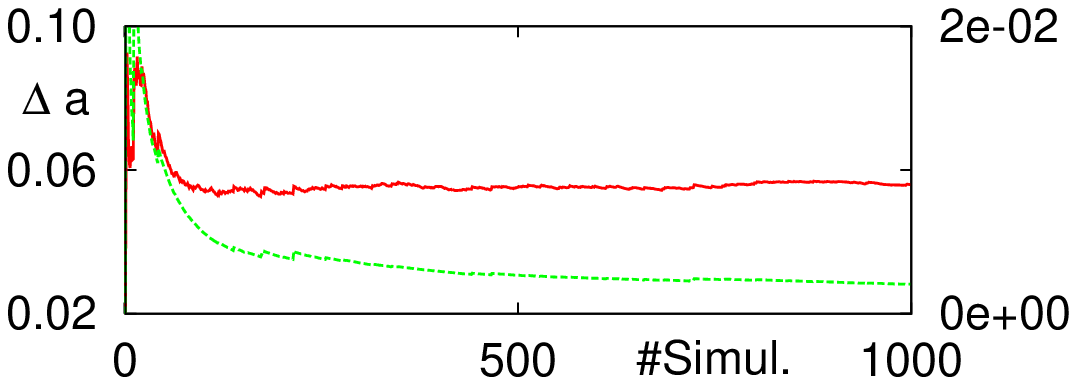}
\includegraphics[width=75mm]{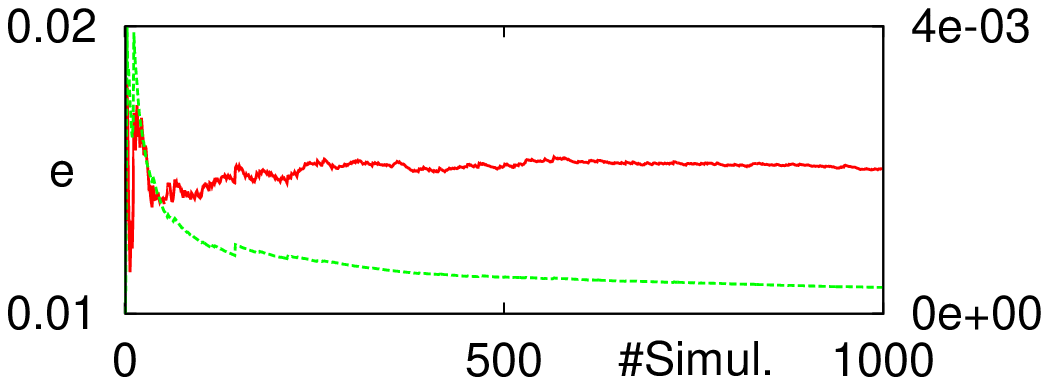}\\
\includegraphics[width=75mm]{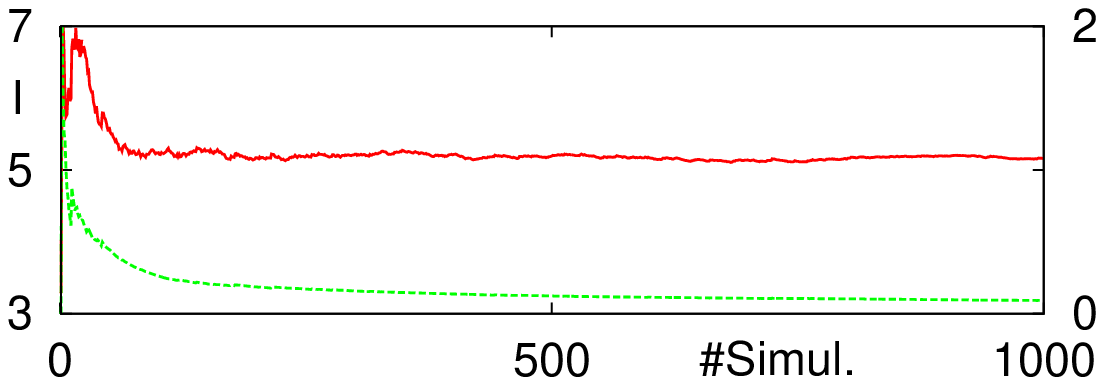}\includegraphics[width=75mm]{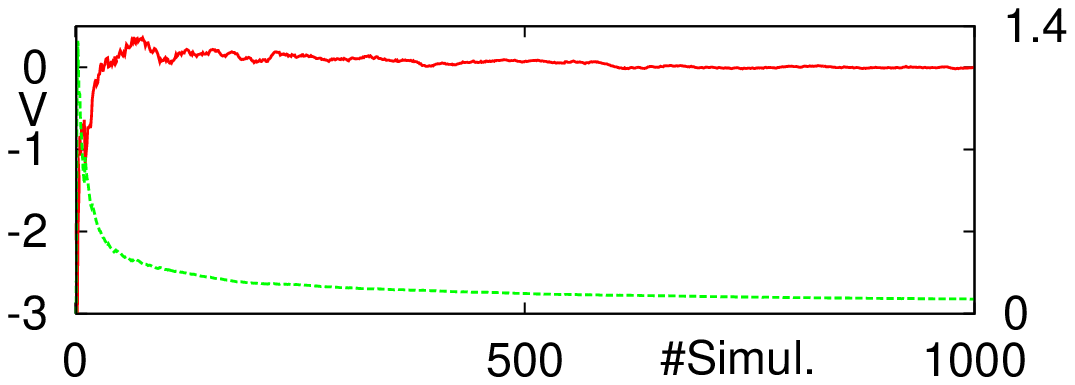}
\end{center}
\label{f3}
\caption{Simulation convergence for a periodic polarity reversal acceleration.  One thousand stochastic acceleration profiles are used with the same parameters as in Figure 1. The solid red lines (left scales) denote the mean values and the dotted green lines (right scales) the standard deviations. The semimajor axis migration rate $\Delta a=a-a_0$ where $a_0=5$ AU is the initial semimajor axis. The mean residual velocity $V$ and its standard deviation are quoted in km\,s$^{-1}$.} 
\end{figure}
\newpage
\begin{figure}
\begin{center}
\includegraphics[width=65mm]{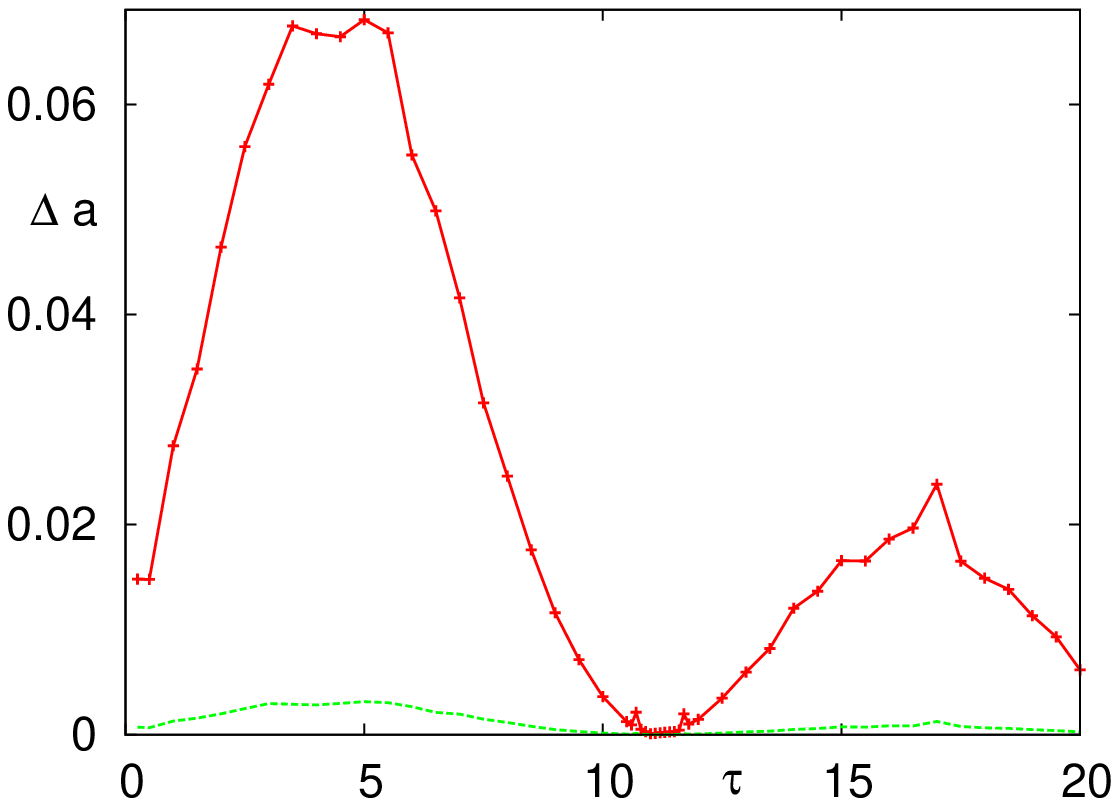}
\includegraphics[width=65mm]{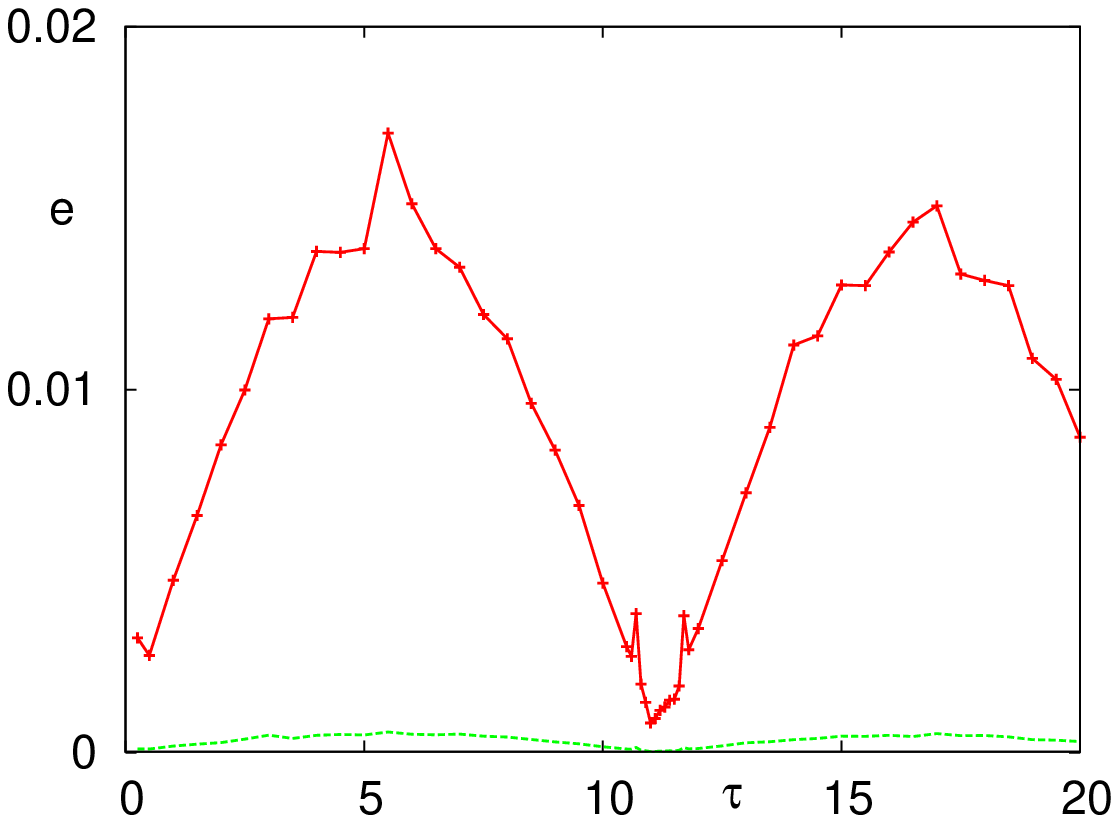}\\
\includegraphics[width=65mm]{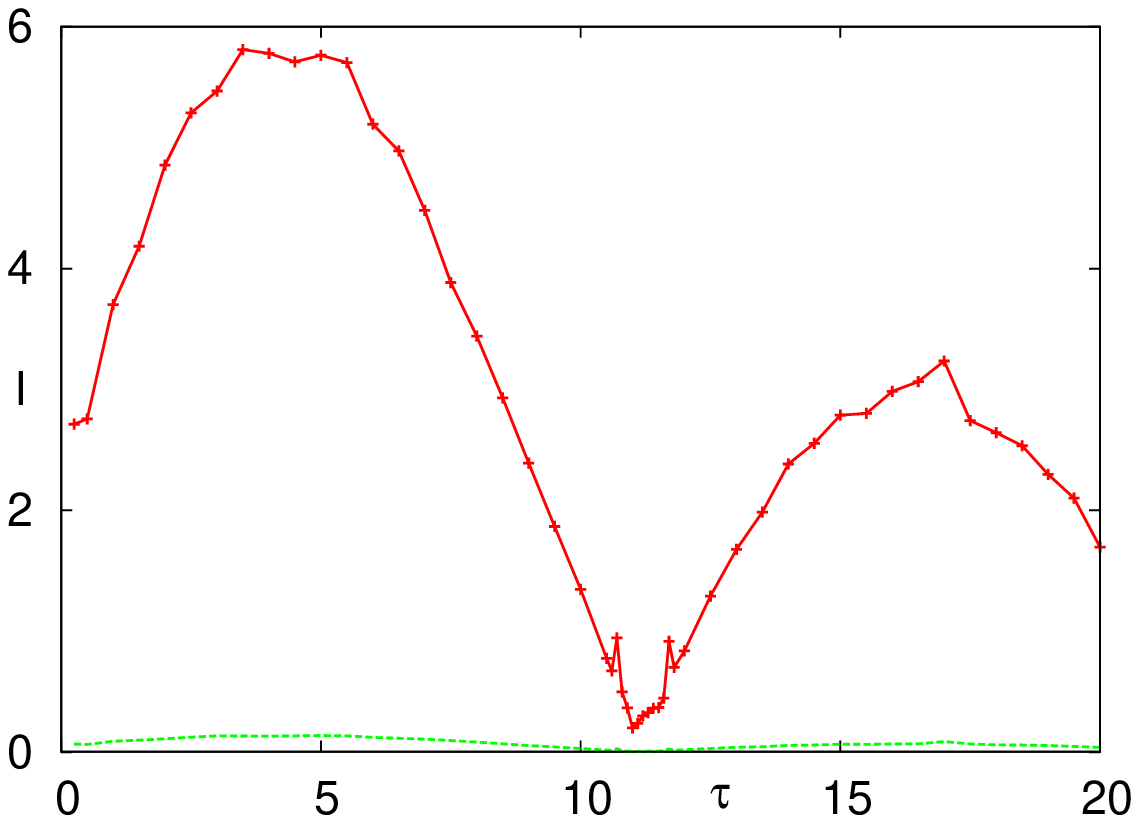}
\includegraphics[width=65mm]{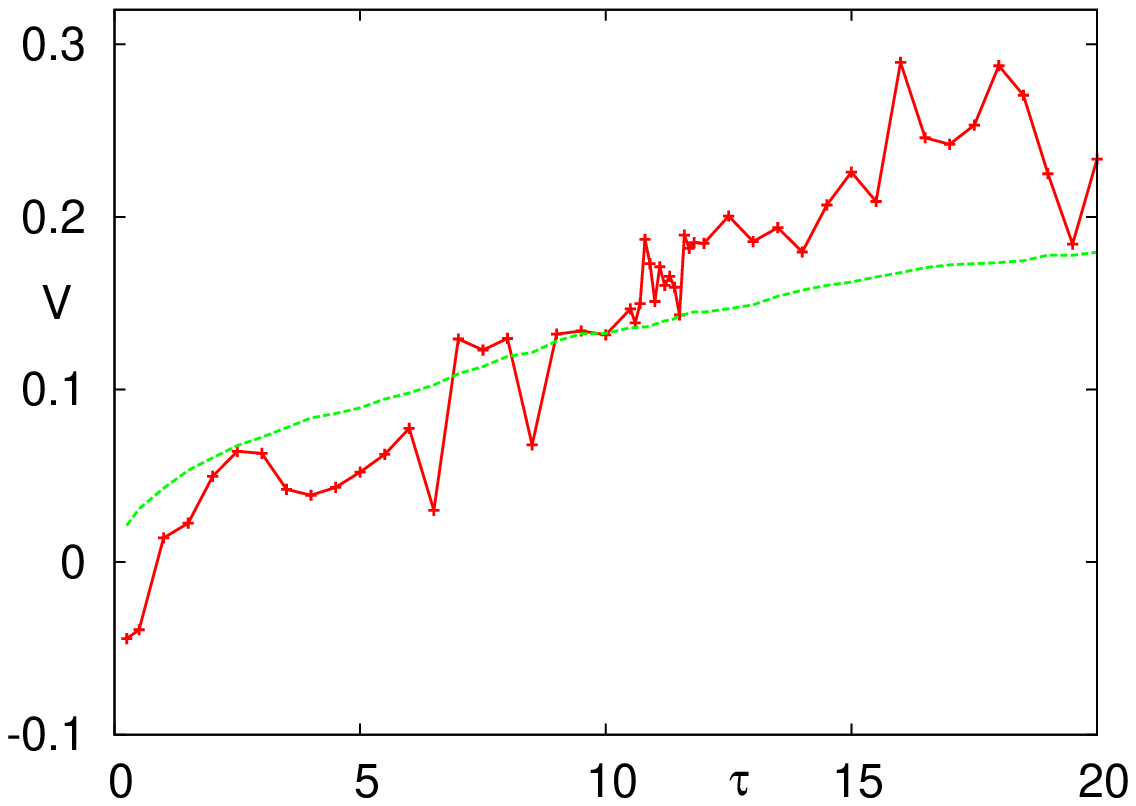}
\end{center}
\label{f4}
\caption{Orbital excitation as a function of variability timescale for a periodic polarity reversal acceleration. The Jupiter-size planet is initially at 5\, AU on a circular orbit.  The variability timescale, $\tau$, is given in years. Each curve point represents the mean of 500 simulations each integrated over $10^6$ years. The solid red lines denote mean values and the dotted  green lines the standard deviations.} 
\end{figure}
\newpage
\begin{figure}
\begin{center}
\includegraphics[width=65mm]{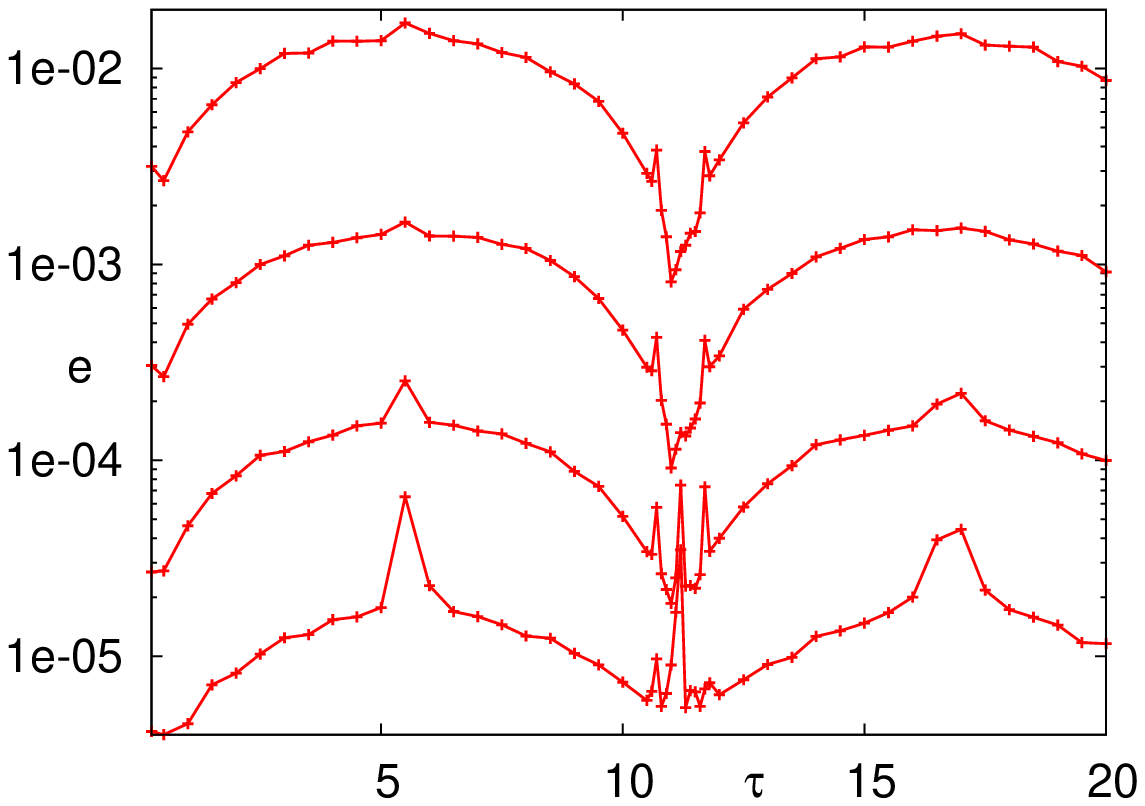} \includegraphics[width=65mm]{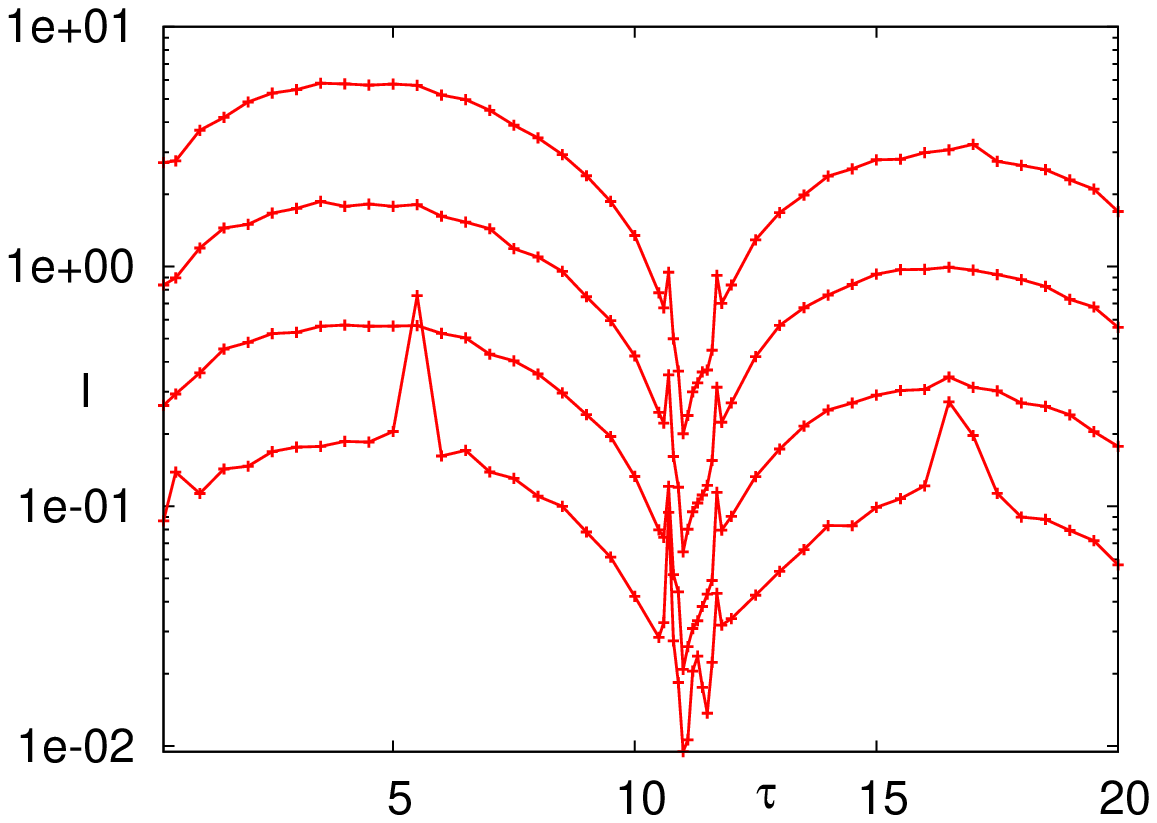}
\end{center}
\label{f5}
\caption{Time evolution of orbital excitation as a function of variability timescale. The parameters are those of Figure 4. In each panel, excitation amplitudes are shown at 4 different times from top to bottom of $10^6$, $10^5$, $10^4$, and $10^3$ years. Standard deviations of $e$ and $I$ are not shown but they have a similar behavior. Each point represents the mean of 500 simulations.} 
\end{figure}
\begin{figure}
\begin{center}
\hspace*{-10mm}\includegraphics[width=60mm]{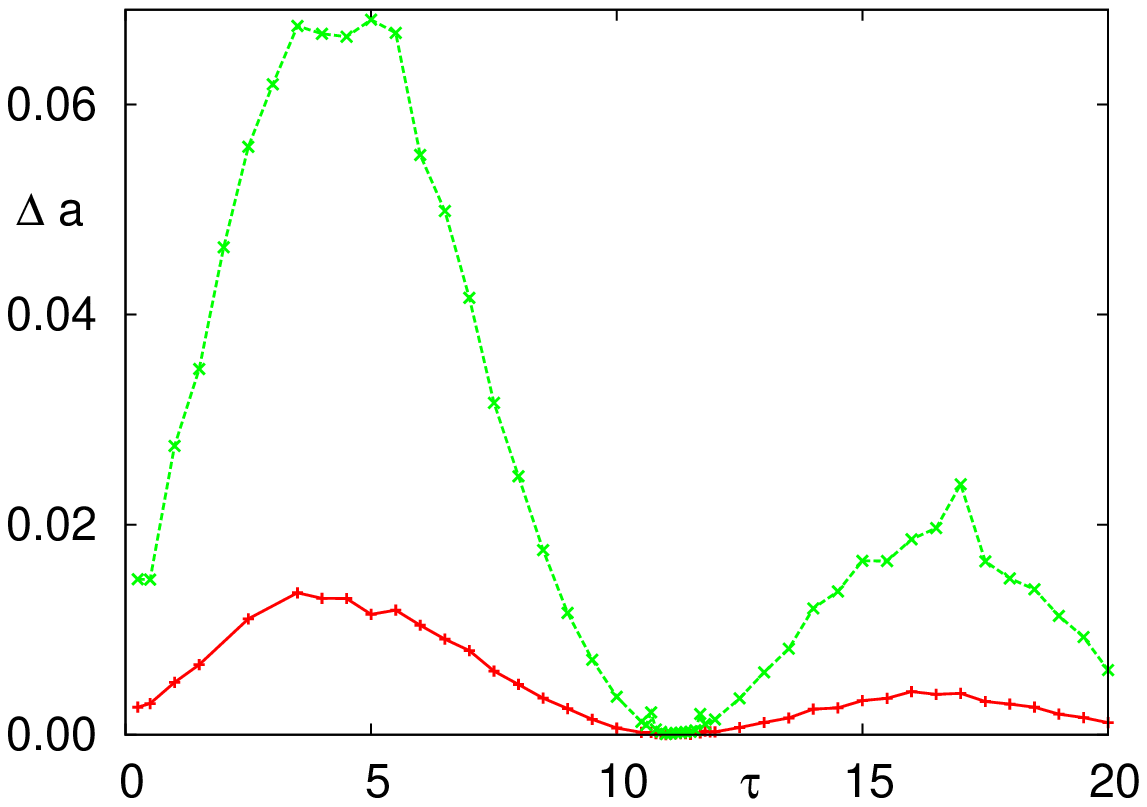}\includegraphics[width=60mm]{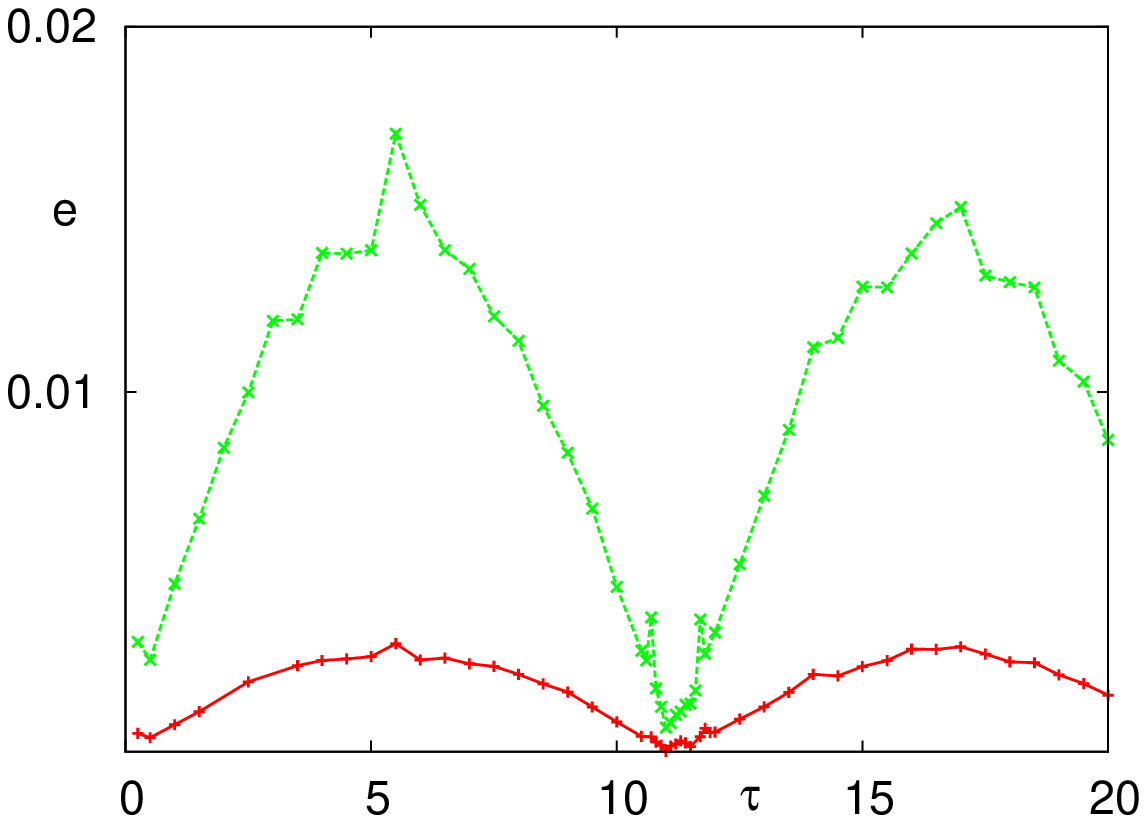}\includegraphics[width=60mm]{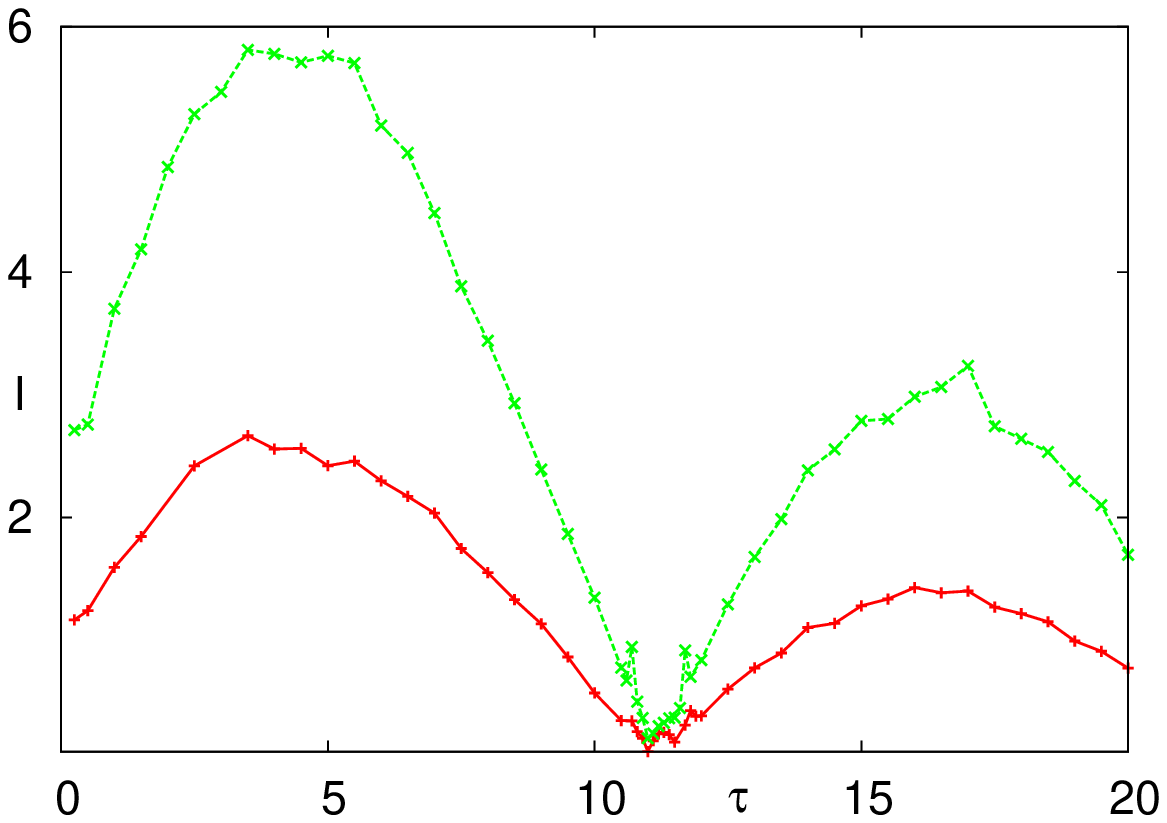}
\end{center}
\label{f6}
\caption{Excitation dependence on acceleration strength. The solid red lines denote the mean excited eccentricity and inclination amplitudes for a standard deviation of $a_{\rm kplr}=300\,$AU and the dotted green line for $a_{\rm kplr}=200\,$AU. Each point represents 500 acceleration profiles.} 
\end{figure}
\newpage
\begin{figure}
\begin{center}
\hspace*{-10mm}\includegraphics[width=60mm]{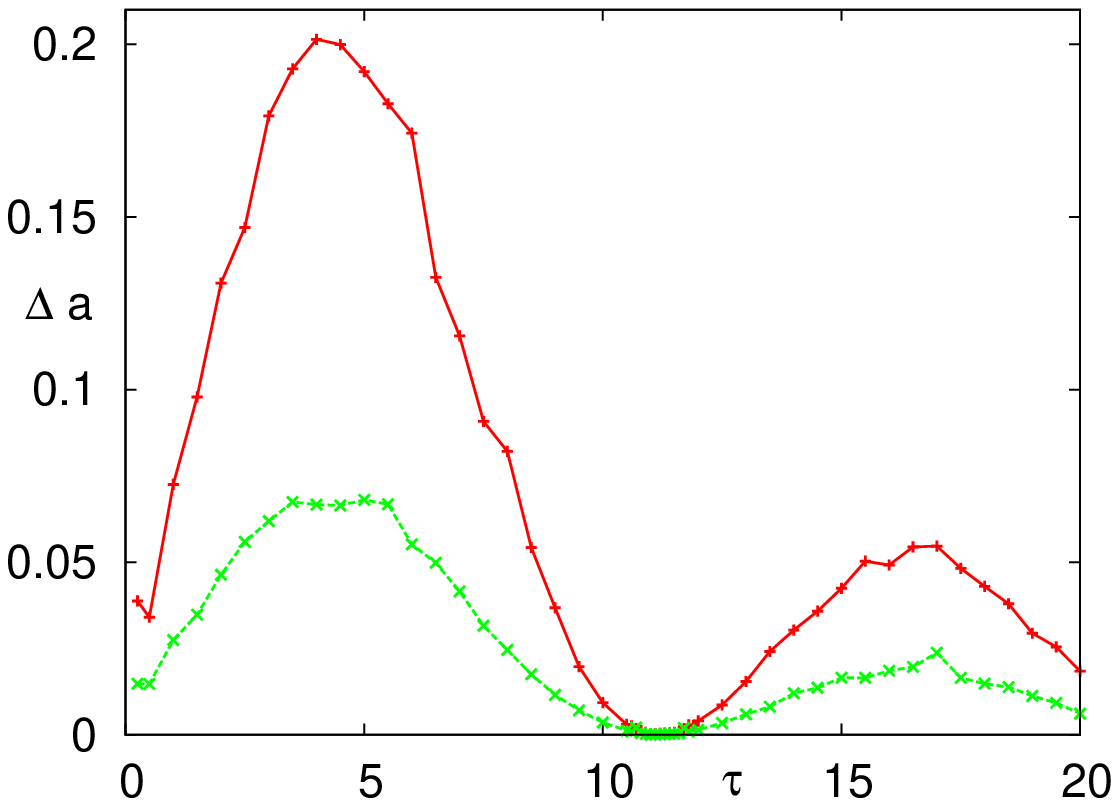}\includegraphics[width=60mm]{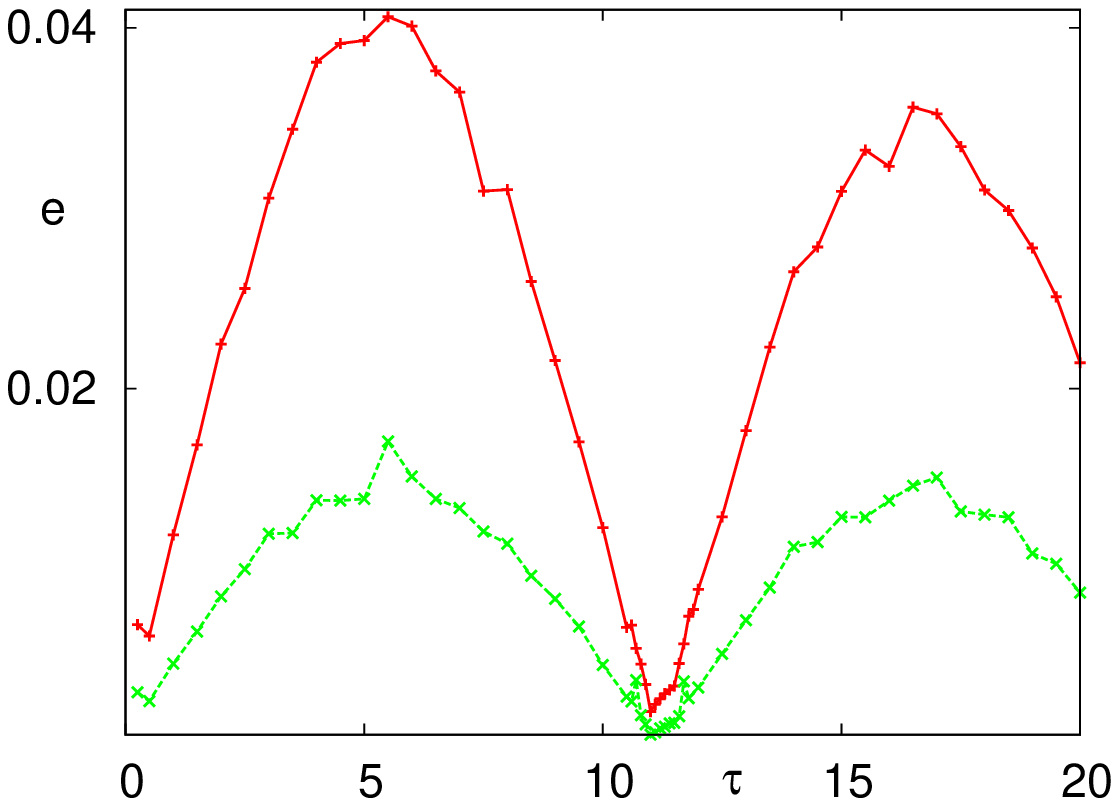}\includegraphics[width=60mm]{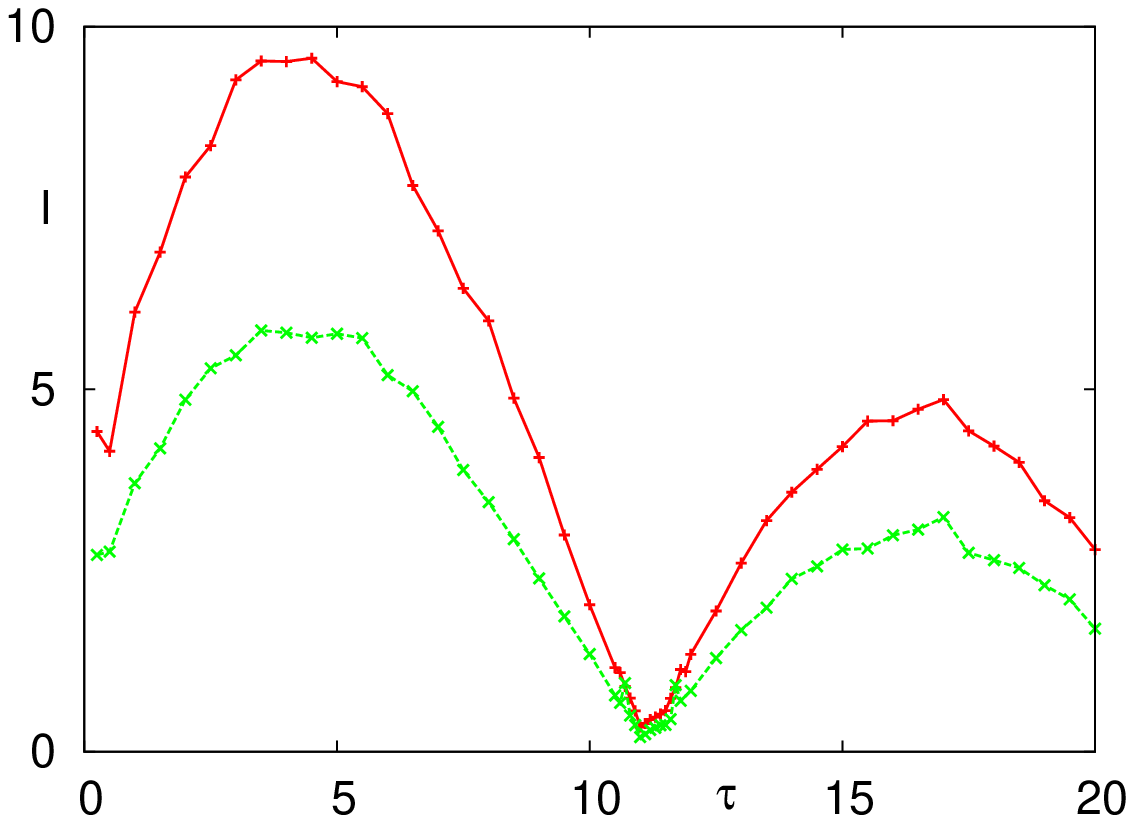}
\end{center}
\label{f7}
\caption{Comparison of random (solid red lines) and periodic (dotted green lines) polarity reversal excitation. The acceleration standard deviation has $a_{\rm kplr}=200\,$AU.  Each curve point represents the mean of 500 simulations each integrated over $10^6$ years. }\end{figure}

\newpage
\begin{figure}
\begin{center}
\hspace*{-10mm}\includegraphics[width=60mm]{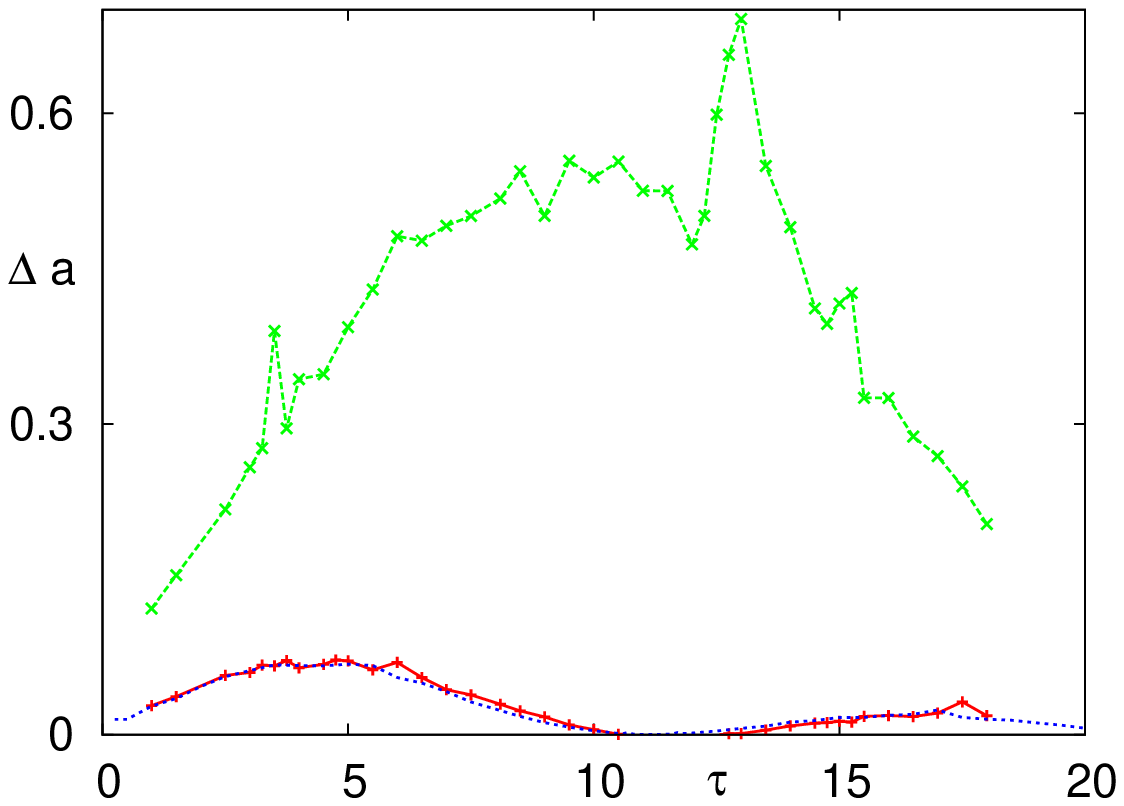}\includegraphics[width=60mm]{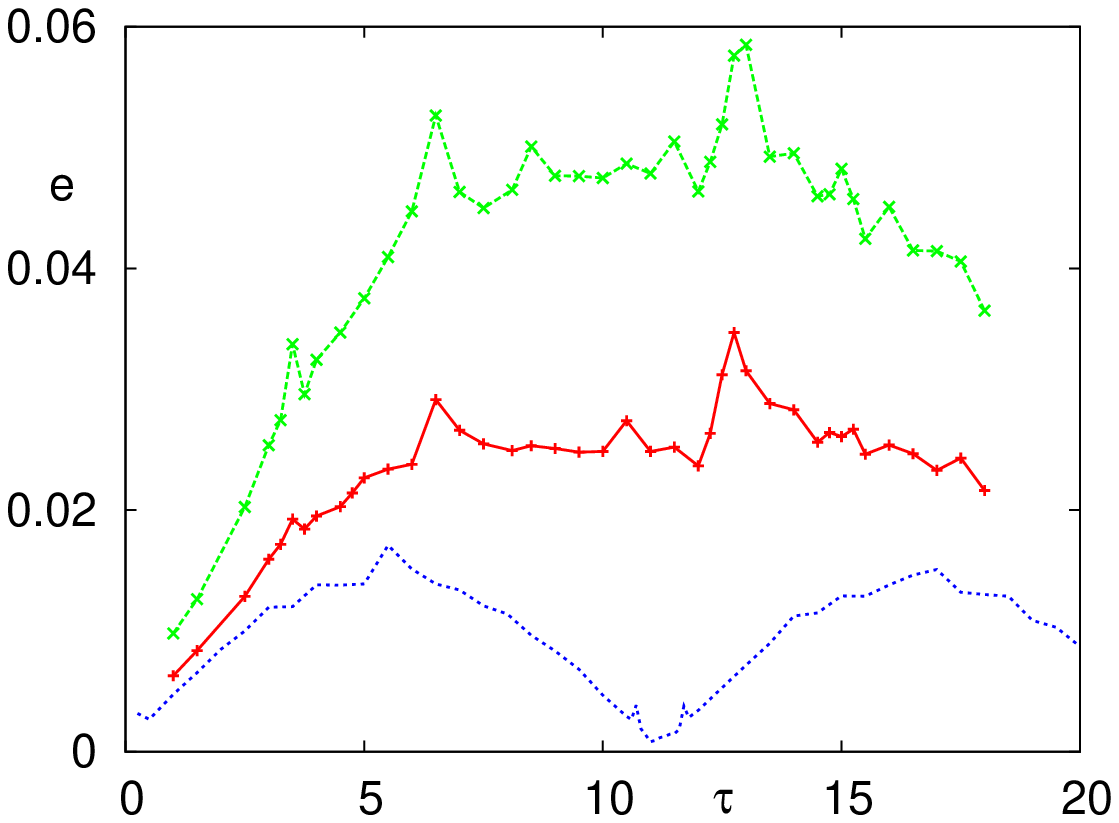}\includegraphics[width=60mm]{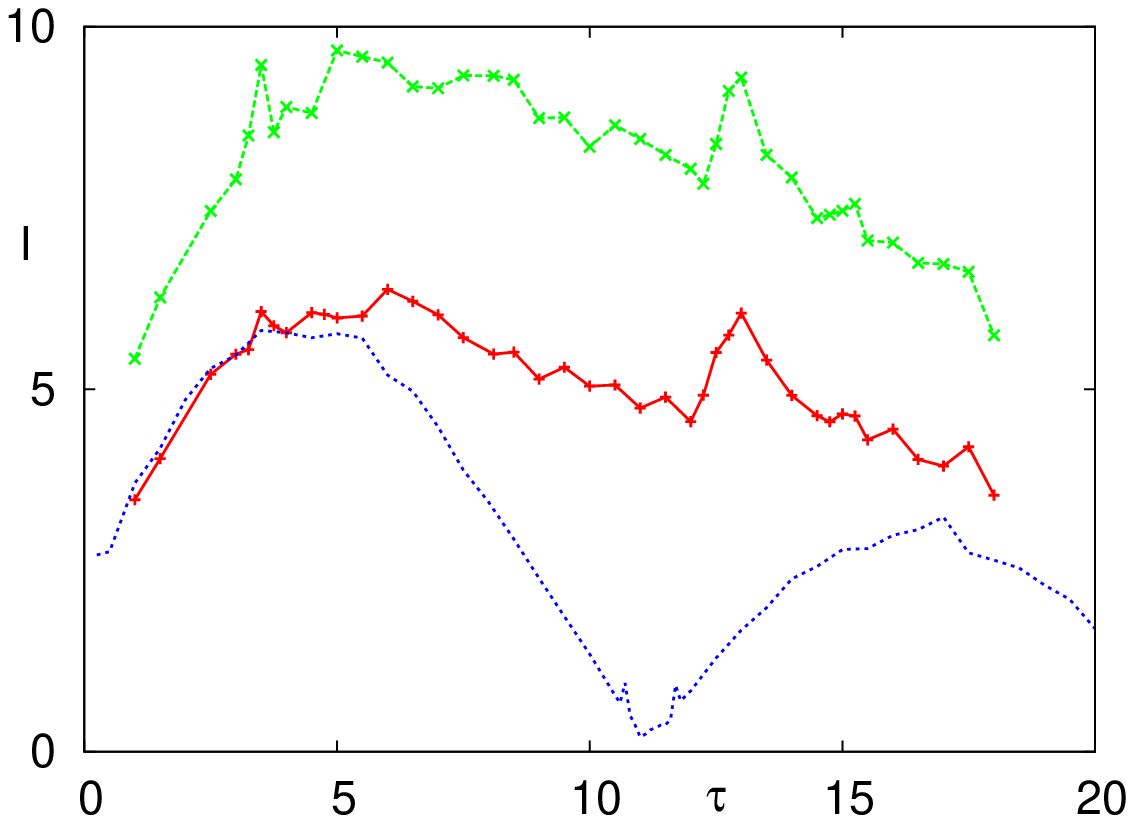}
\end{center}
\label{f8}
\caption{Influence of mutual planet interactions on orbital excitation. The planets initially are at 5\,AU (Jupiter, solid red line) and at 8.5\,AU (Saturn, dashed green line) both on circular orbits.  The acceleration standard deviation has $a_{\rm kplr}=200\,$AU. The blue dotted lines represent the mean excitation amplitudes of Jupiter alone shown in Fig. 4.  Each curve point represents the mean of 500 simulations each integrated over $10^6$ years. } 
\end{figure}
\newpage
\begin{figure}
\begin{center}
\hspace*{-10mm}\includegraphics[width=60mm]{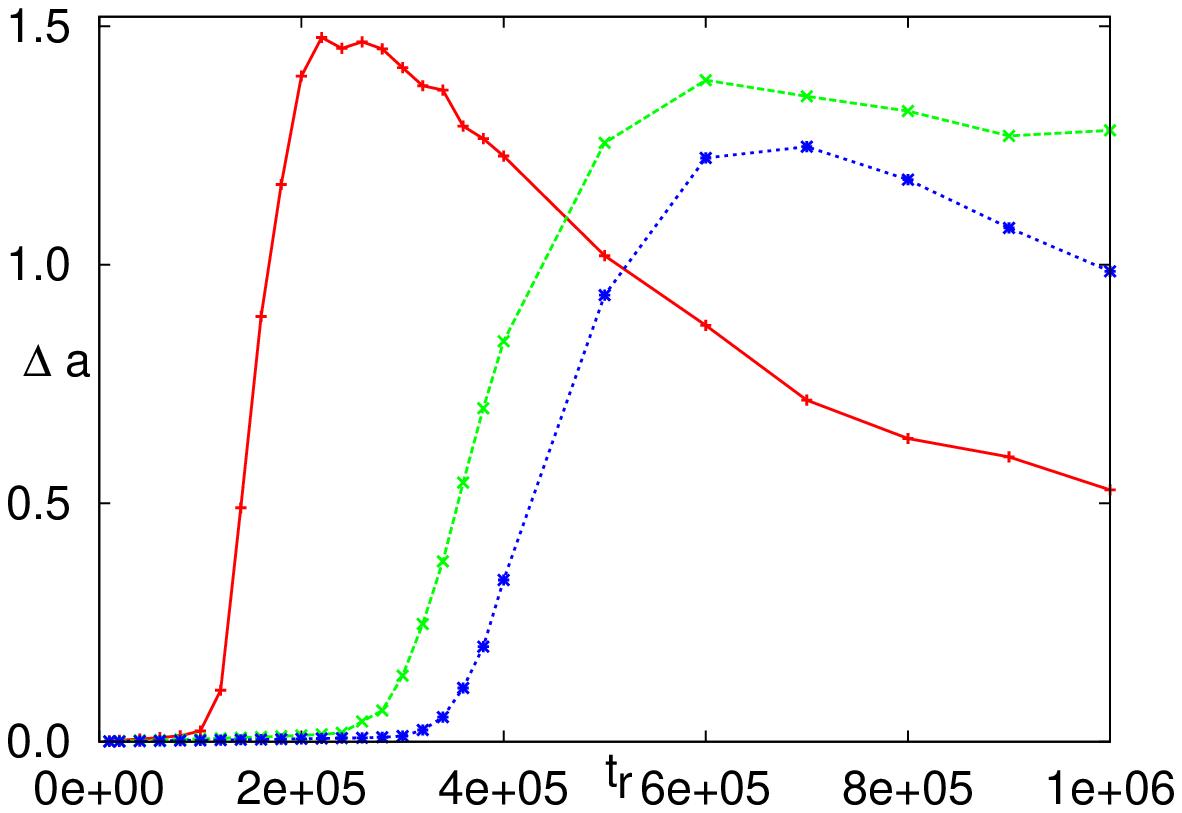}\includegraphics[width=60mm]{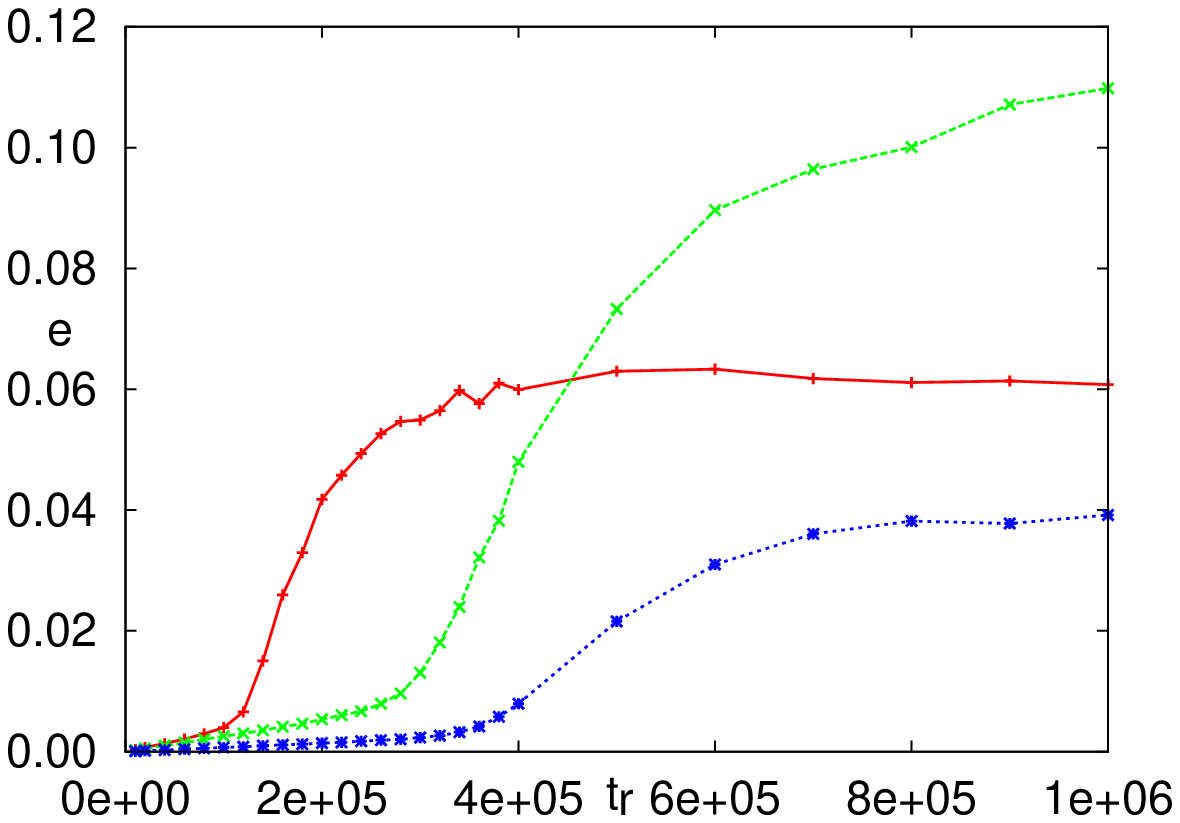}\includegraphics[width=60mm]{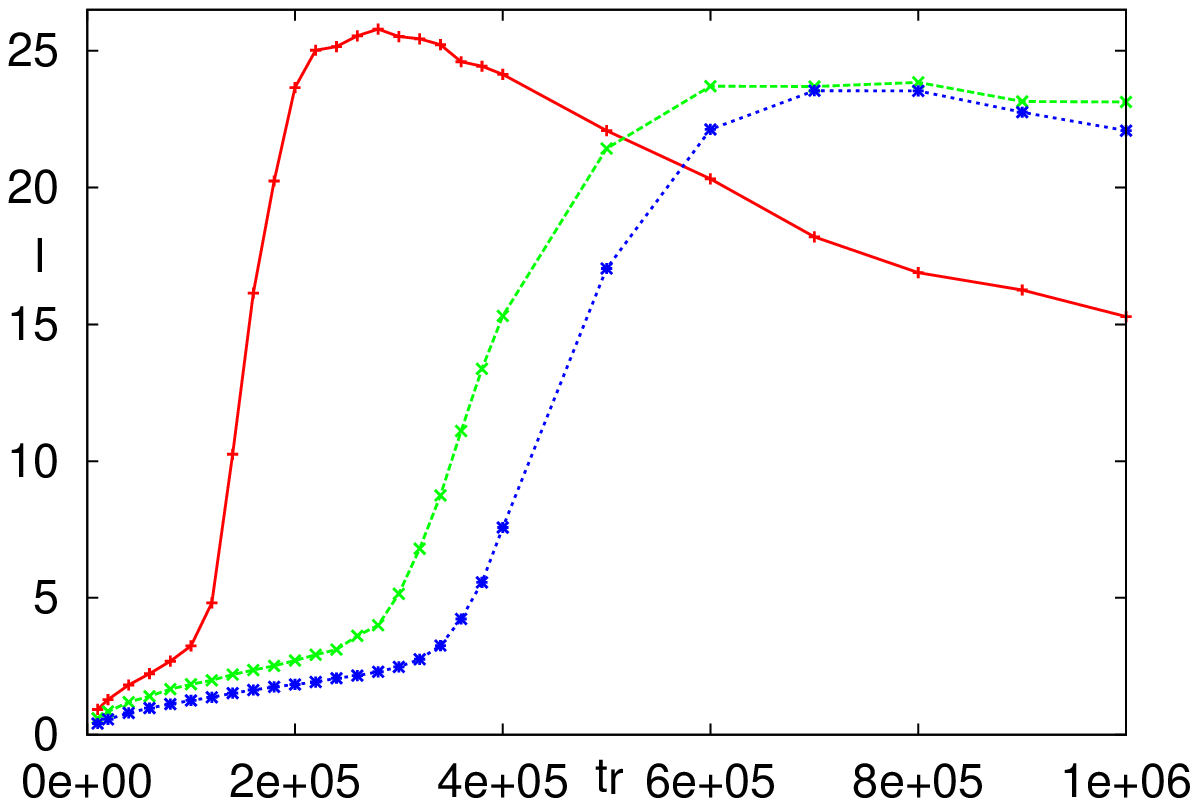}\\
\hspace*{-10mm}\includegraphics[width=60mm]{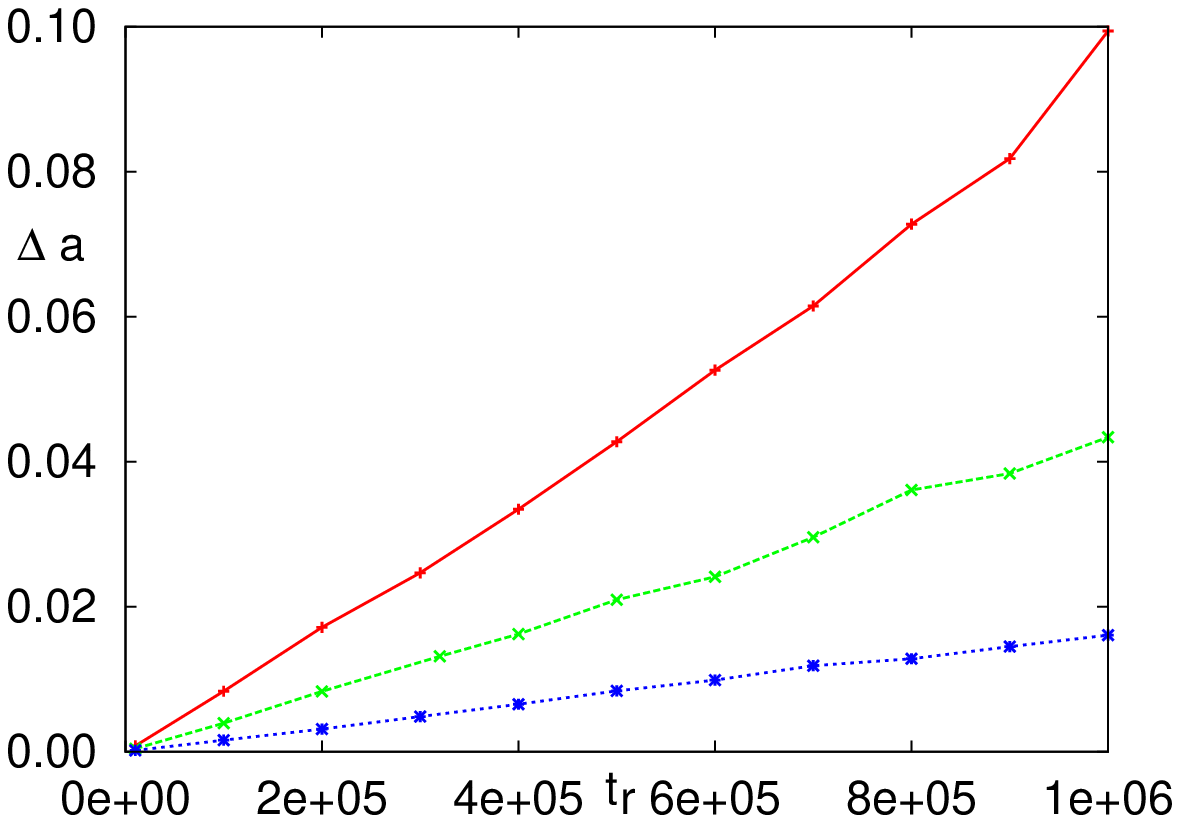}\includegraphics[width=60mm]{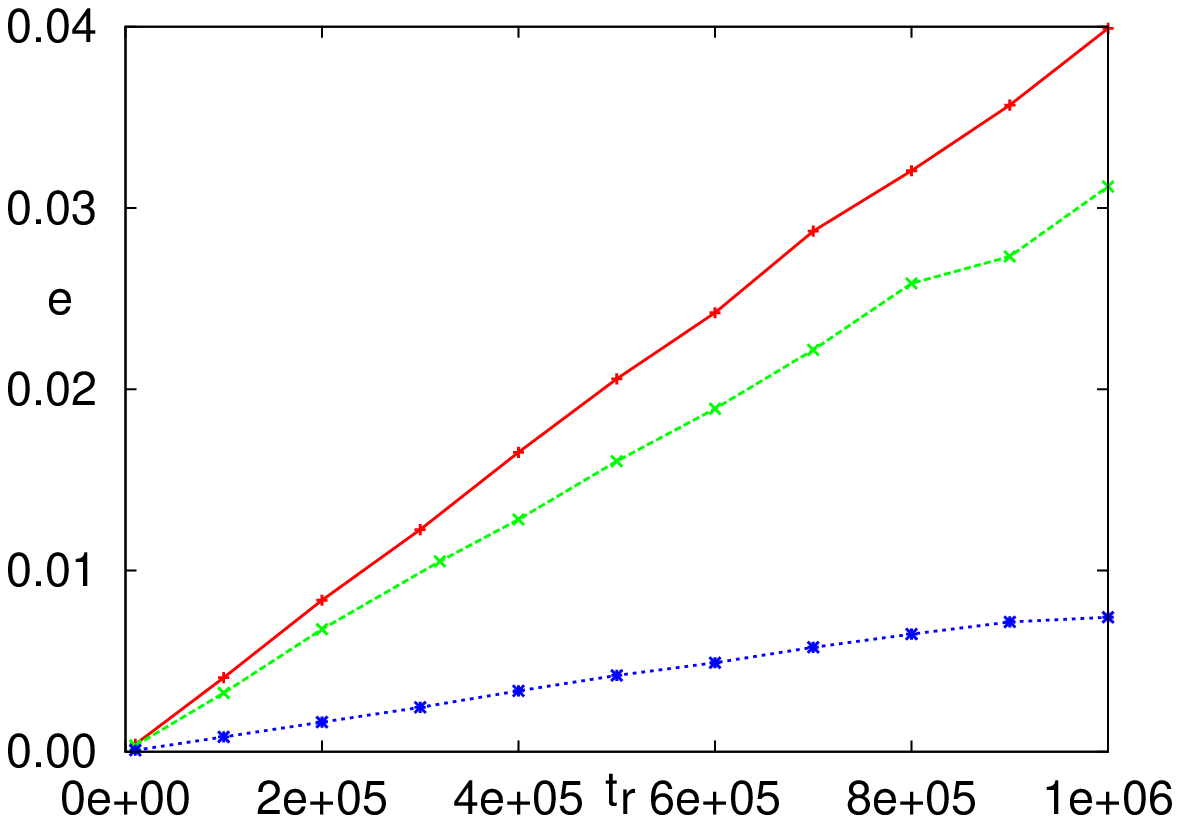}\includegraphics[width=60mm]{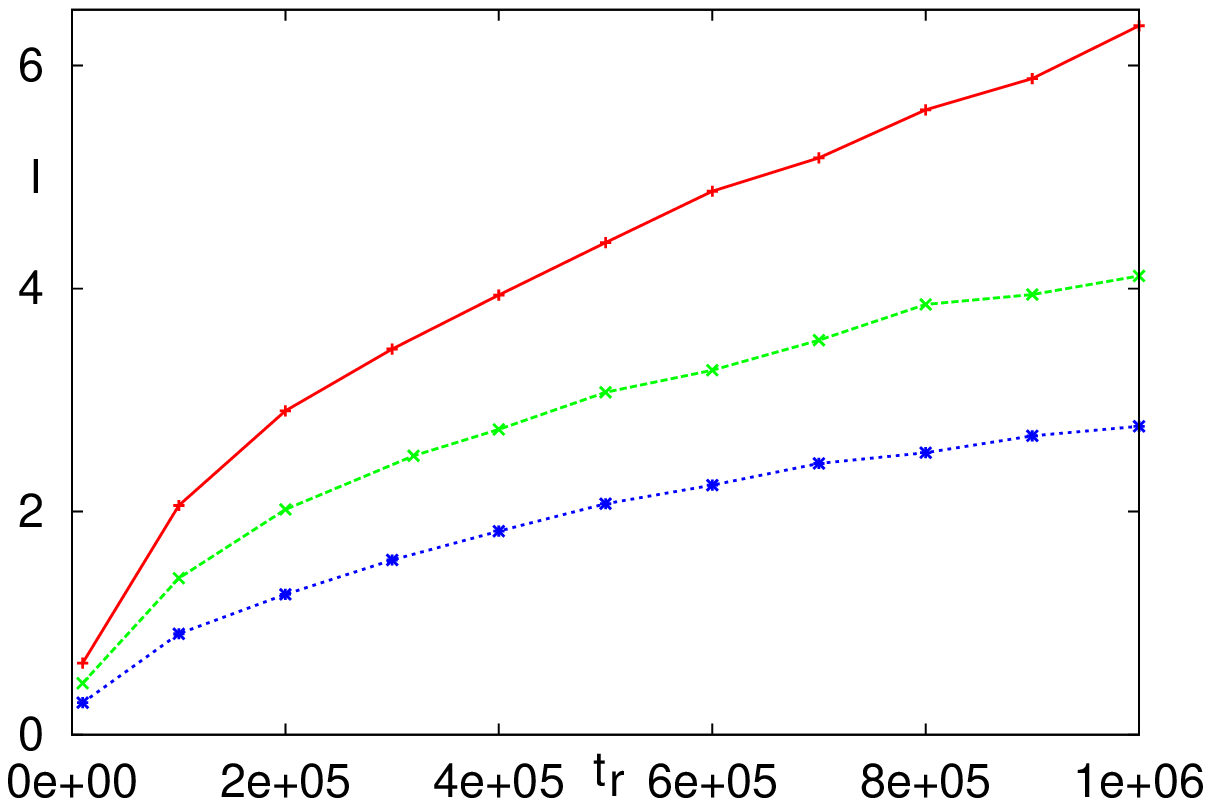}
\end{center}
\label{f9}
\caption{Resonance crossing for periodic  and random polarity reversal excitation. The upper (lower) row corresponds to periodic (random) reversals. The Jupiter-size planet at 5 AU is subjected to stochastic momentum loss excitation with a time-dependent variability timescale $\tau$ that takes $t_r$ years to reach 11 years starting from 6 months.  The solid red lines correspond to $a_{\rm kplr}=200$\,AU and a power law increase of $\tau$ with exponent 0.5. The dashed green lines correspond to $a_{\rm kplr}=200$\,AU  and exponential increase of $\tau$ with a relaxation time of $t_r/5$. The dotted blue lines correspond to $a_{\rm kplr}=300$\,AU and  a power law increase of $\tau$ with exponent 0.5. Each curve point represents the mean of 2000 simulations.}
\end{figure}
\newpage
\begin{figure}
\begin{center}
\hspace*{-10mm}\includegraphics[width=60mm]{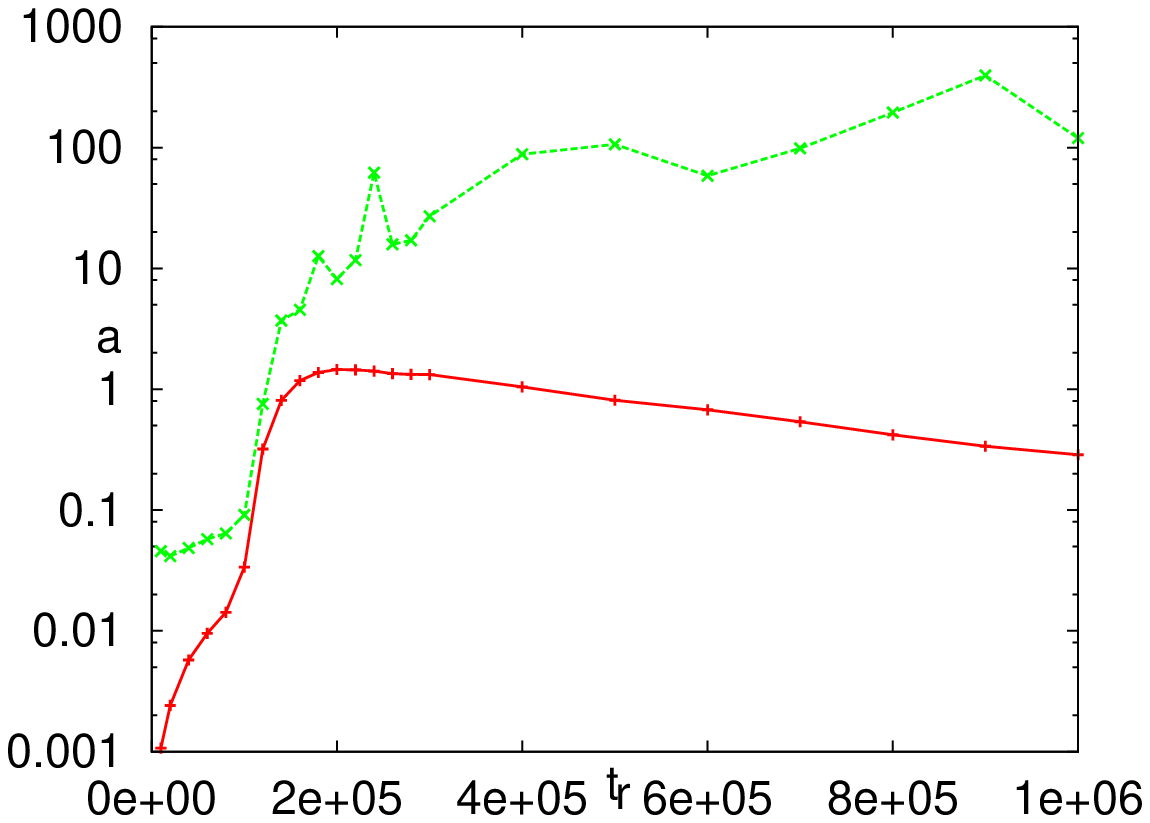}\includegraphics[width=60mm]{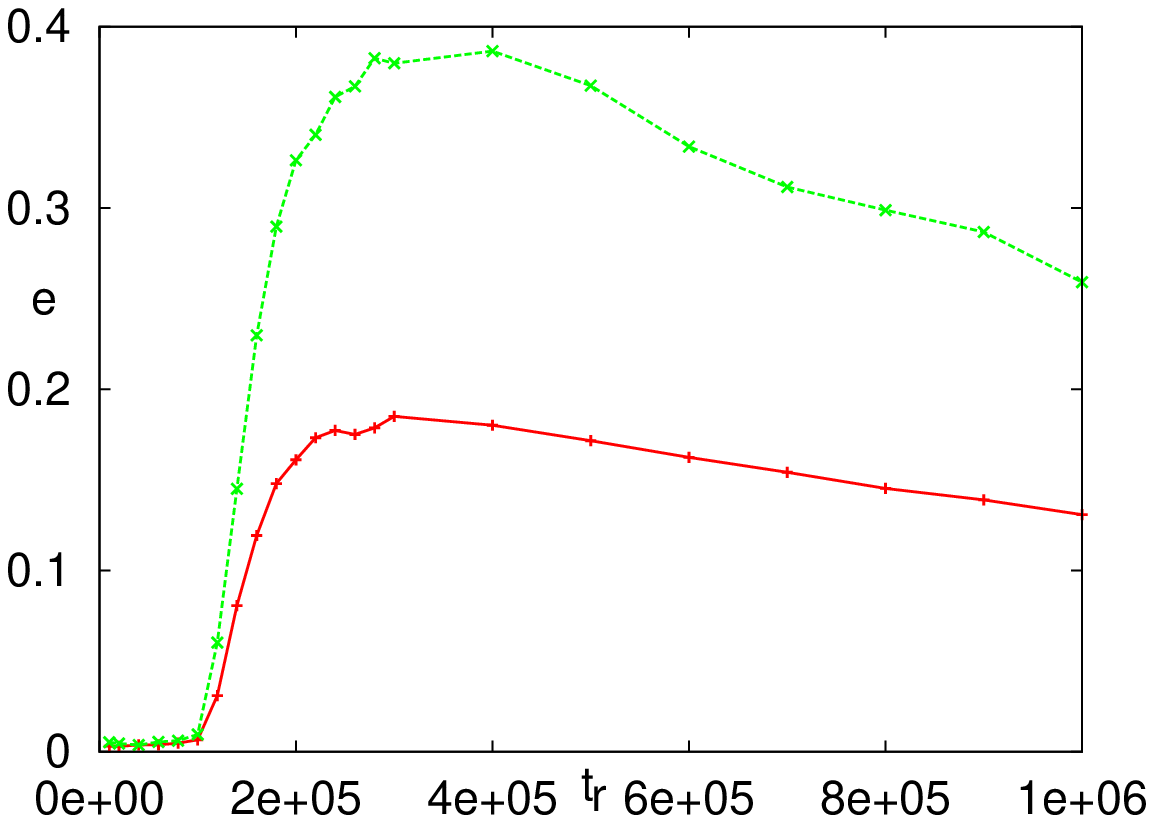}\includegraphics[width=60mm]{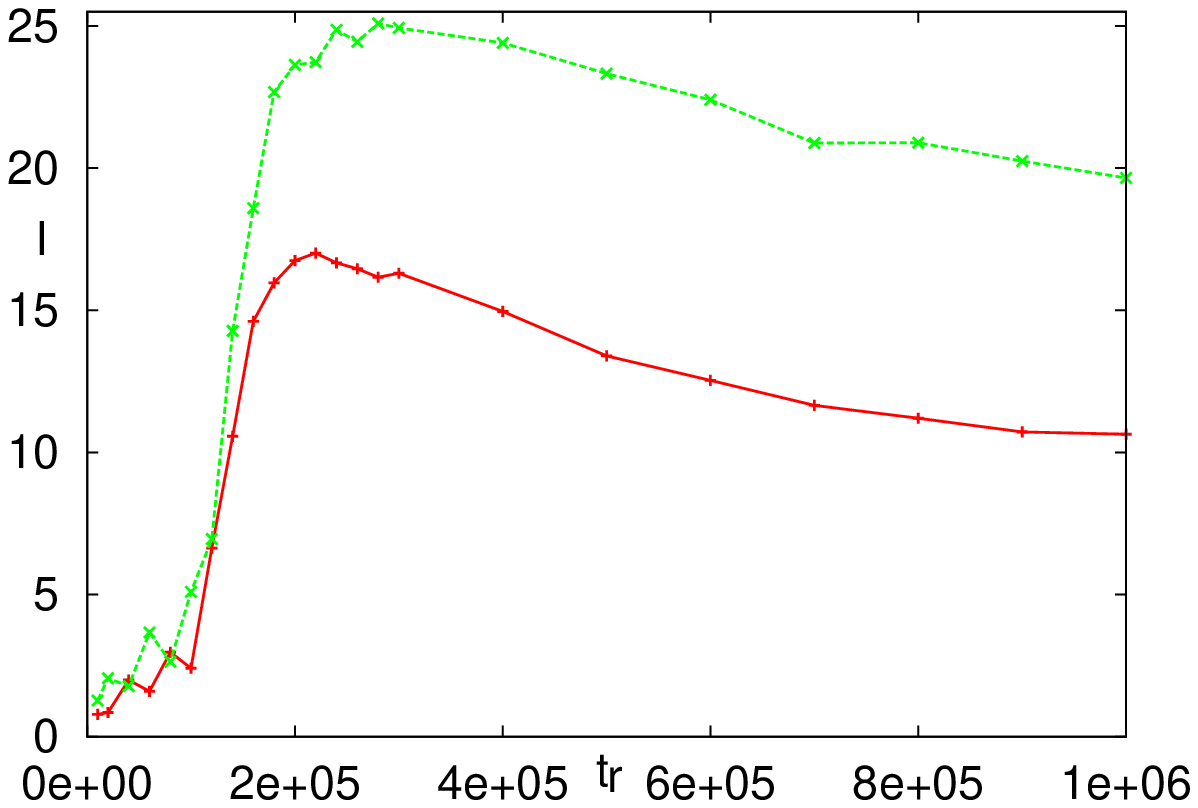}\\
\hspace*{-10mm}\includegraphics[width=60mm]{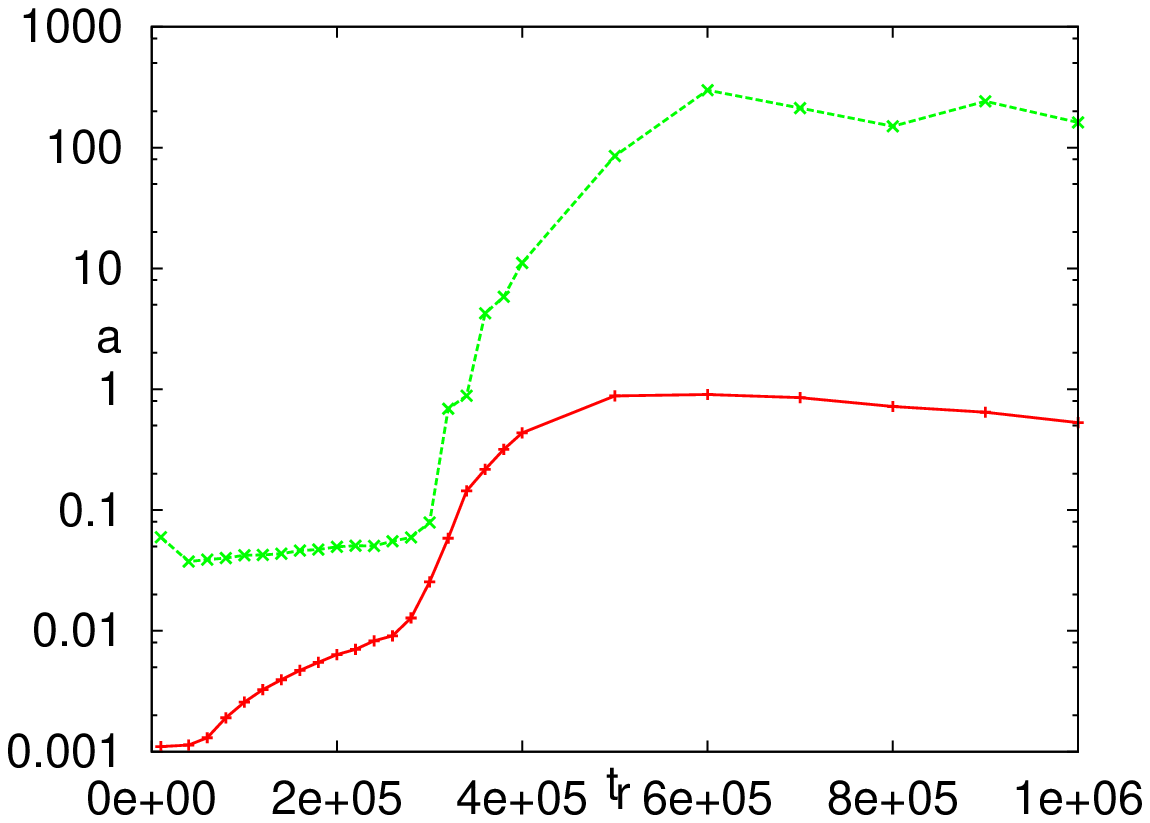}\includegraphics[width=60mm]{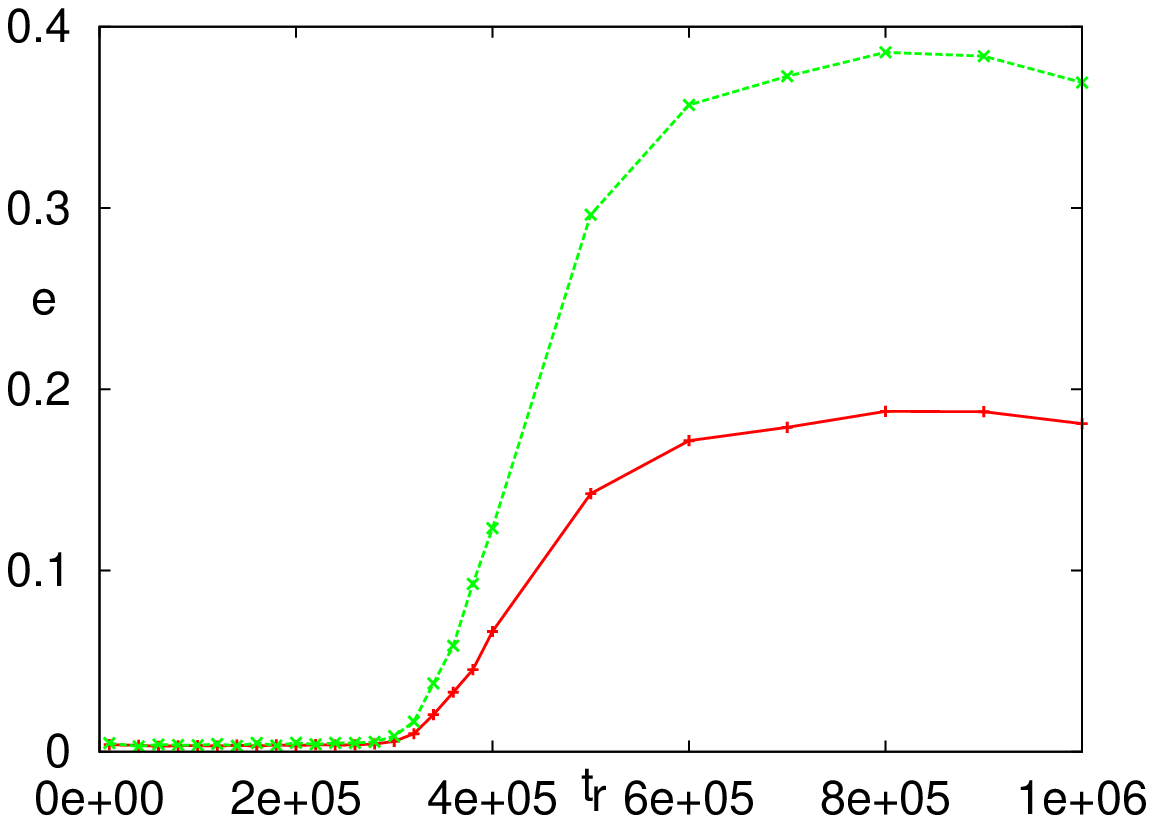}\includegraphics[width=60mm]{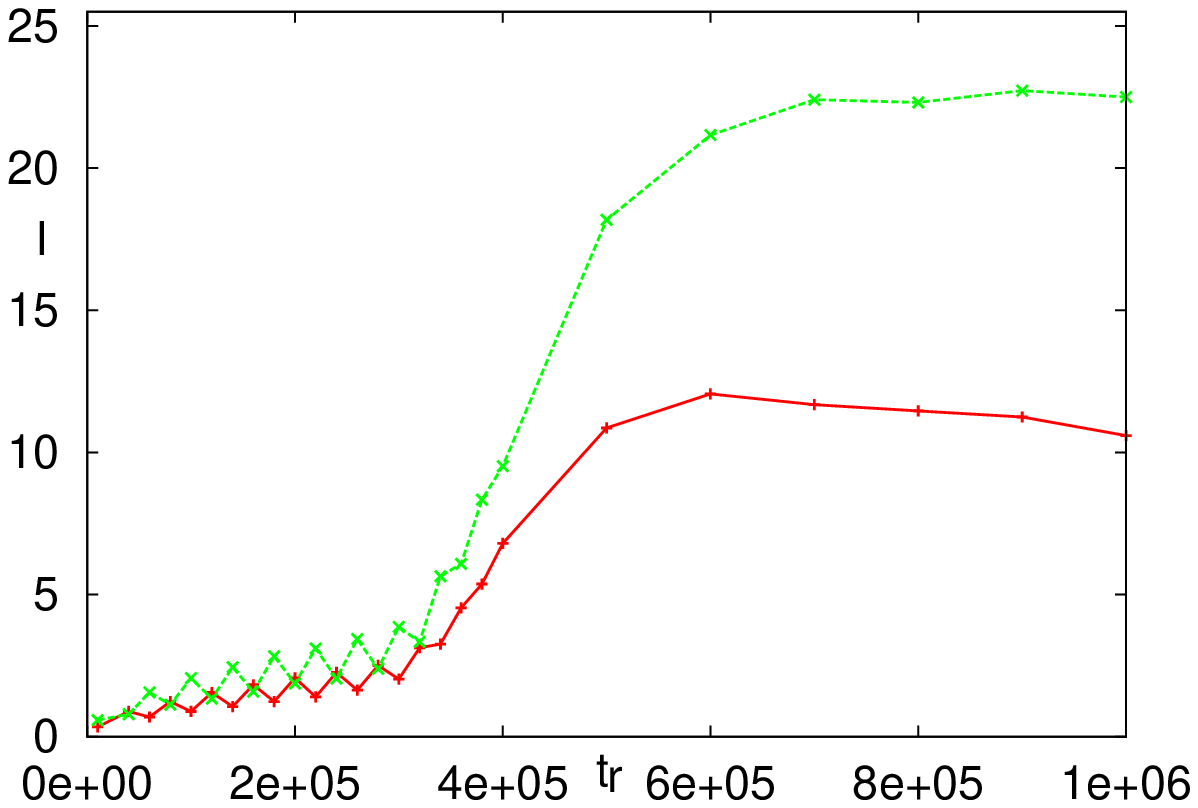}\\
\hspace*{-10mm}\includegraphics[width=60mm]{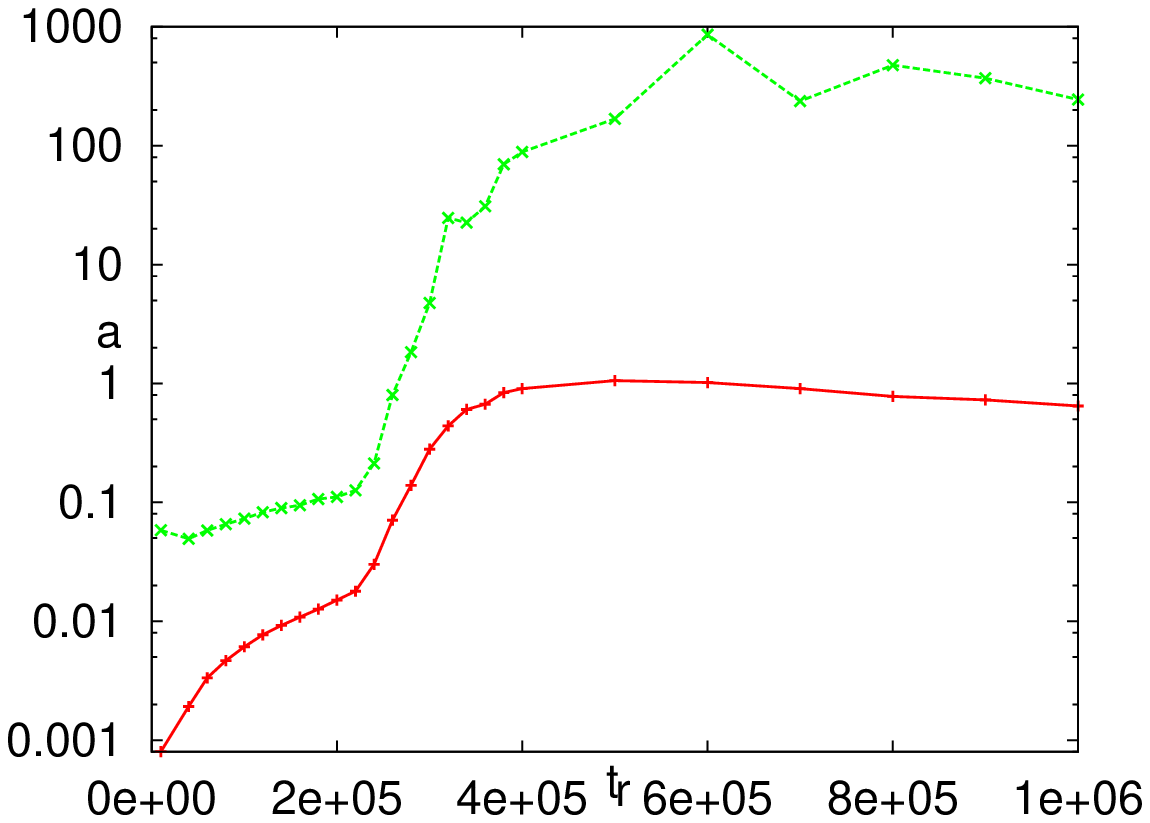}\includegraphics[width=60mm]{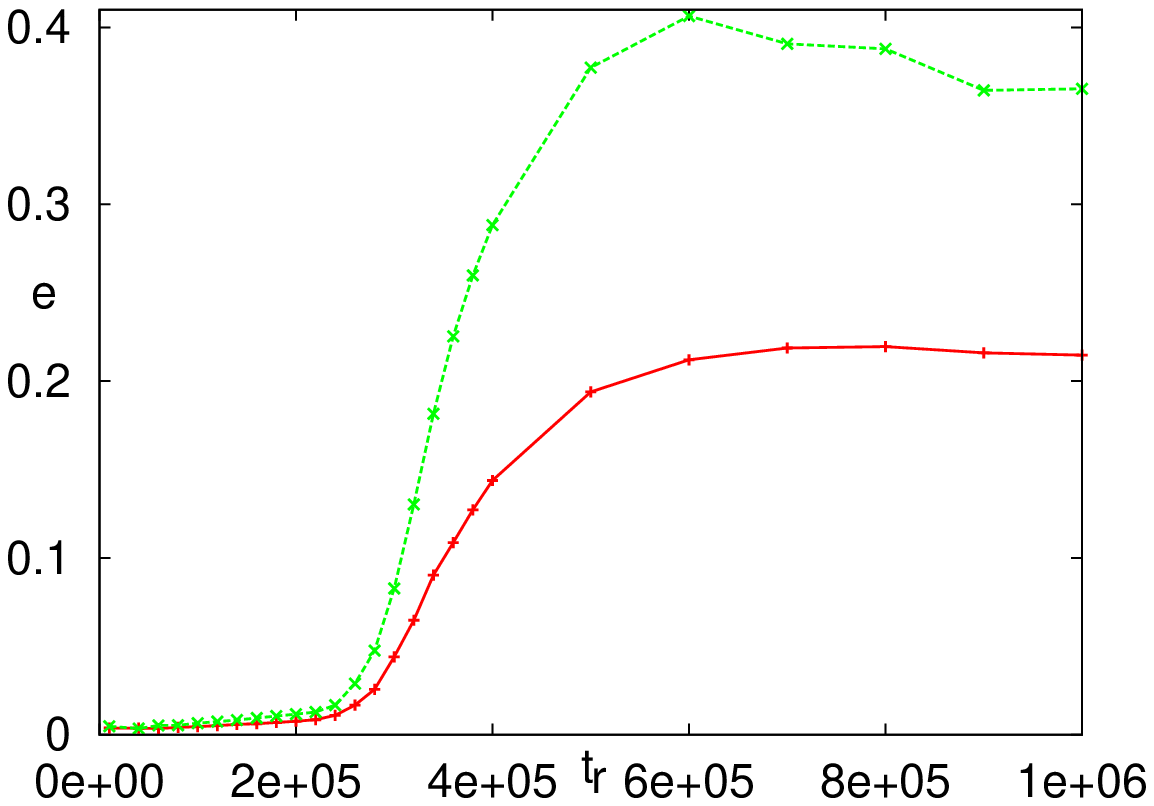}\includegraphics[width=60mm]{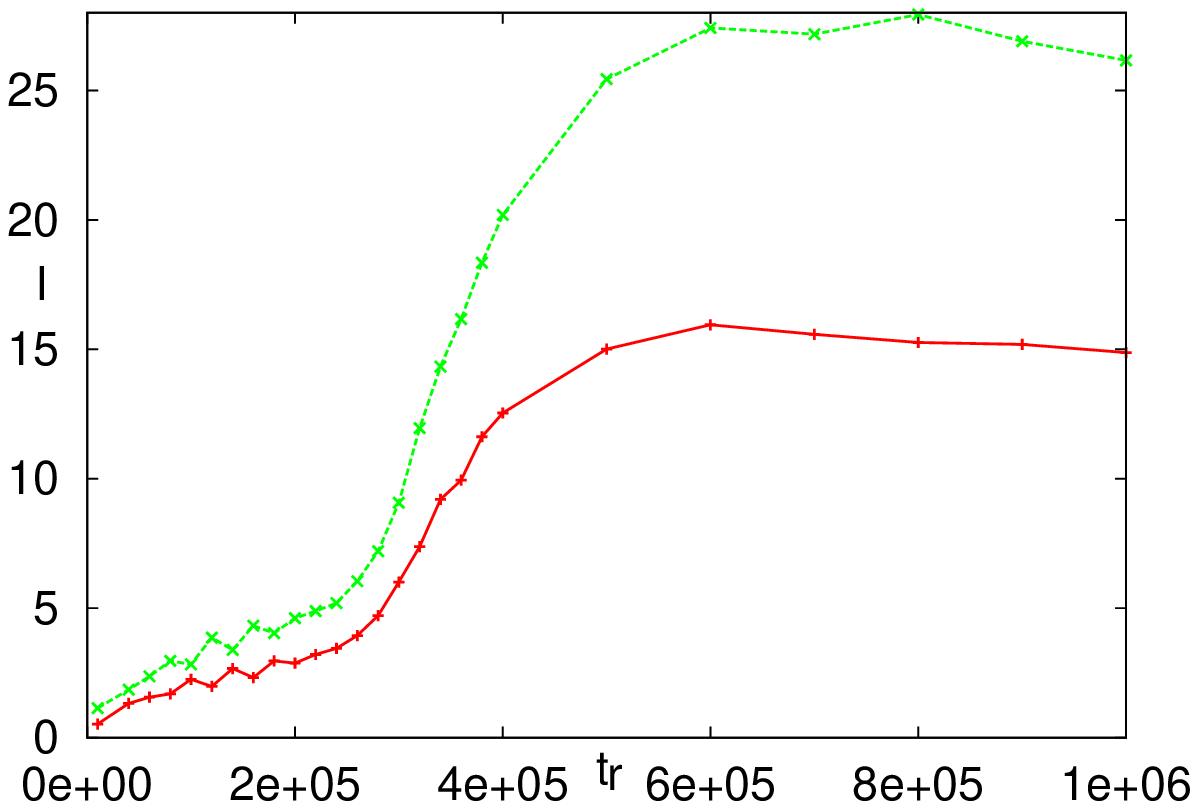}\\
\hspace*{-10mm}\includegraphics[width=60mm]{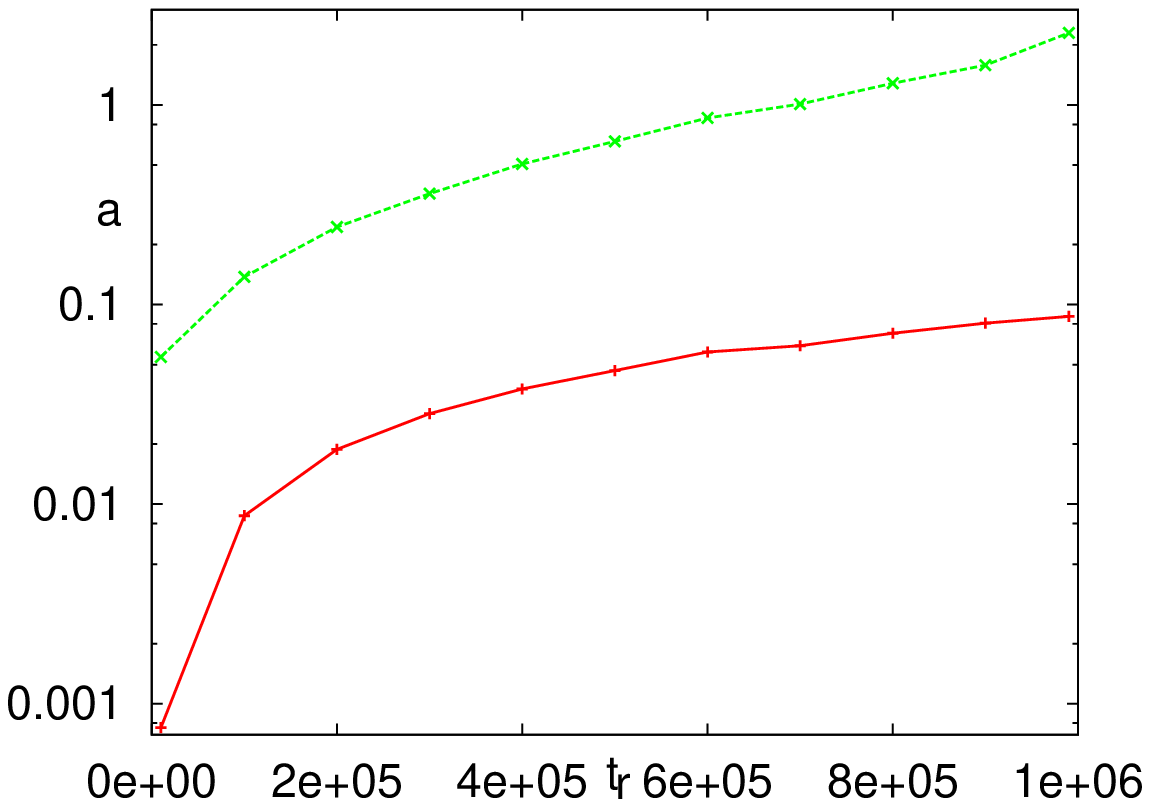}\includegraphics[width=60mm]{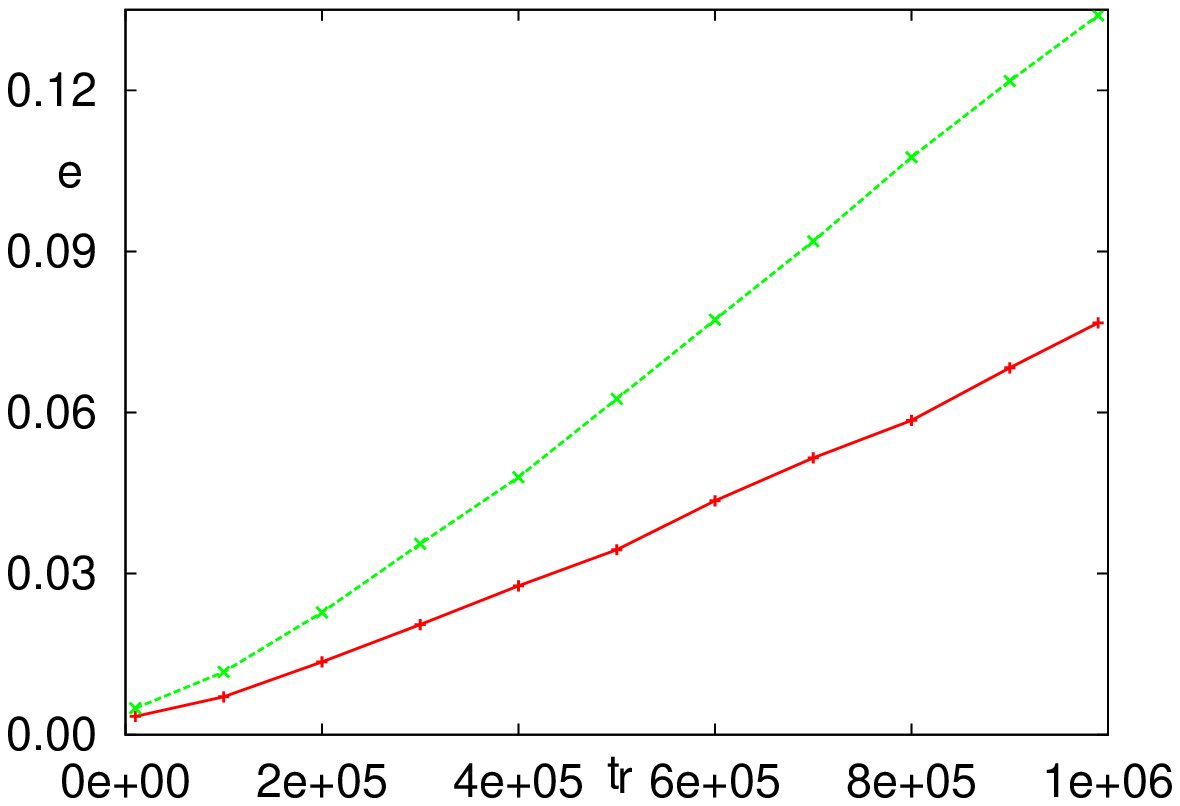}\includegraphics[width=60mm]{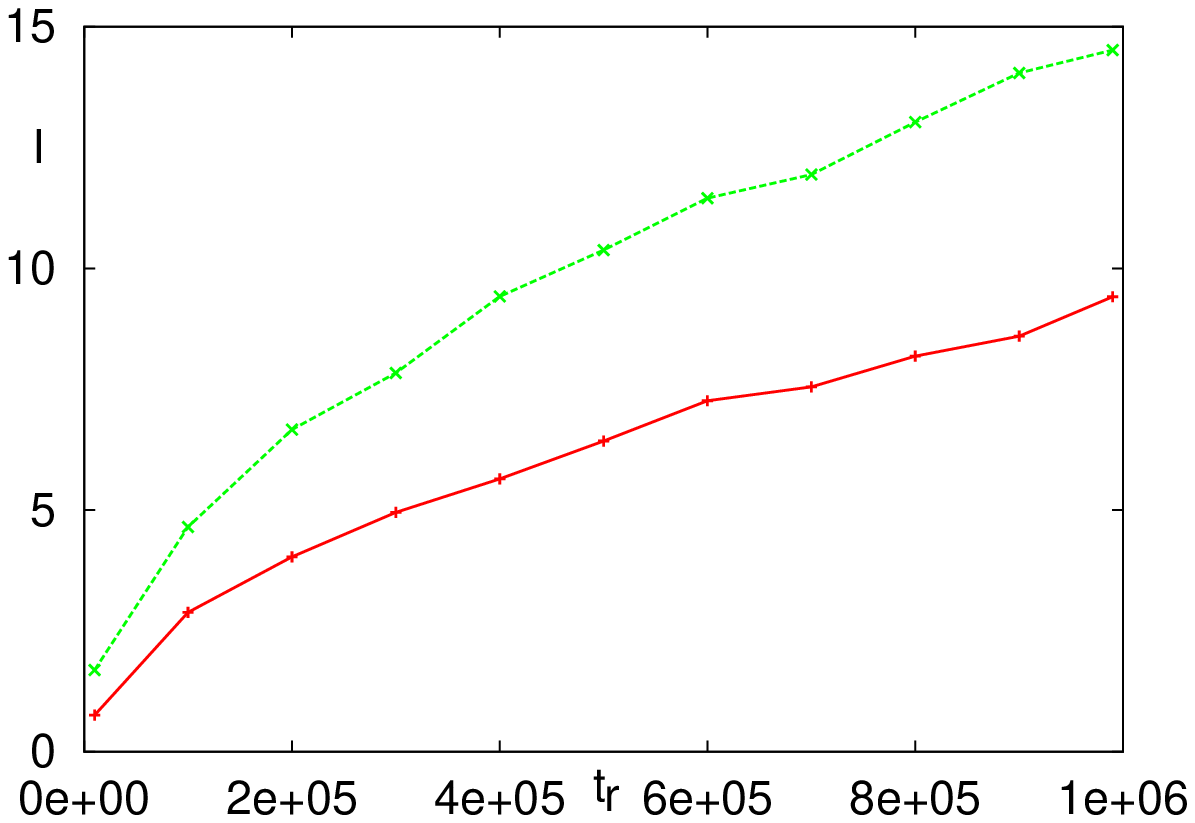}
\end{center}
\label{f10}
\caption{Resonance crossing for two interacting planets. Jupiter (solid red lines) and Saturn (dashed green lines) are initially at 5\,AU and 8.5\,AU and are subjected to stochastic momentum loss excitation with  time-dependent variability timescale $\tau$ that takes $t_r$ years to reach 11 years starting from 6 months. The first and second rows corresponds to periodic reversal with a power law increase of $\tau$ of exponent 0.5 and respectively $a_{\rm kplr}=200$\,AU and  $a_{\rm kplr}=300$\,AU. The third row corresponds to  periodic reversal with exponential increase of $\tau$ and a relaxation time of $t_r/5$ and $a_{\rm kplr}=200$\,AU. The bottom row corresponds to random reversal with $a_{\rm kplr}=200$\,AU and  a power law increase of $\tau$ of exponent 0.5. Each curve point represents the mean of 2000 simulations.} 
\end{figure}

\end{document}